\documentclass[useAMS,usenatbib]{mn2e}

\usepackage{natbib, aas_macros}
\usepackage{ulem}
\usepackage[dvips]{graphicx}
\usepackage{wrapfig}
\graphicspath{{./figs/}}
\usepackage{color}
\usepackage{amsmath}
\usepackage{amssymb}
\newcommand{\Msun}{\rm{ M_{\odot}}}

\newcommand{\Nion}{\dot{N}_{\rm ion}^{\gamma}}
\newcommand{\nion}{\dot{n}_{\rm ion}^{\gamma}}

\newcommand{\La}{L_{\rm Ly\alpha}}

\newcommand{\Mstar}{M_{\rm{star}}}

\newcommand{\lya}{\rm {Ly{\alpha}}}

\newcommand{\Mh}{M_{\rm h}}
\newcommand{\Msunyr}{\rm M_{\odot}~ yr^{-1}}

\newcommand{\kms}{\rm km\;s^{-1}}

\newcommand{\Qhii}{Q_{\rm HII}}

\newcommand{\ergs}{{\rm erg~s^{-1}}}

\newcommand{\nh}{n_{\rm H}}

\newcommand{\rhii}{R_{\rm HII}}
\newcommand{\fescuv}{f_{\rm esc, UV}}
\newcommand{\fesca}{f_{\rm esc, Ly\alpha}}
\newcommand{\fescion}{f_{\rm esc, ion}}
\newcommand{\ergscm}{erg~s^{-1}~cm^{-2}}
\newcommand{\Flya}{F_{\rm Ly\alpha}}
\newcommand{\Llya}{L_{\rm Ly\alpha}}
\newcommand{\nlae}{n_{\rm LAE}}
\newcommand{\Mgas}{M_{\rm gas}}
\newcommand{\fboost}{f_{\rm boost}^{\rm HII}}

\title[Lyman-alpha Emitting Galaxies and Ionized Bubbles] 
{Modelling of Lyman-alpha Emitting Galaxies and Ionized Bubbles at the Era of Reionization}
\author[Yajima et al.]
{Hidenobu Yajima$^{1, 2}$\thanks{E-mail: yajima@astr.tohoku.ac.jp (HY)}, Kazuyuki Sugimura$^{2}$, 
Kenji Hasegawa$^{3}$
\\
$^{1}$ Frontier Research Institute for Interdisciplinary Sciences, Tohoku University, Sendai 980-8578, Japan\\
$^{2}$ Astronomical Institute, Tohoku University, Sendai 980-8578, Japan\\
$^{3}$ 
Graduate School of Science, Nagoya University, Furo-cho, Chikusa-ku, Nagoya, Aichi 464-8602, Japan\\
}

\begin{document}

\date{Accepted ?; Received ??; in original form ???}

\pagerange{\pageref{firstpage}--\pageref{lastpage}} \pubyear{2008}

\maketitle

\label{firstpage}

%
%
\begin{abstract}
Understanding $\lya$ emitting galaxies (LAEs) can be a key to reveal
cosmic reionization and galaxy formation in the early Universe.  
Based on halo merger trees and $\lya$ radiation transfer calculations, 
we model redshift evolution of LAEs and their observational properties at $z \ge 6$.
We consider ionized bubbles associated with individual LAEs and IGM transmission of $\lya$ photons.
We find that $\lya$ luminosity tightly correlates
with halo mass and stellar mass, while the relation with star
formation rate has a large dispersion.
Comparing our models with the observed luminosity function by Konno et al., 
we suggest that LAEs at $z \sim 7$ have
galactic wind of $V_{\rm out} \gtrsim 100~\rm km\, s^{-1}$ and H{\sc i}
column density of $N_{\rm HI} \gtrsim 10^{20}~\rm cm^{-2}$.
Number density of bright LAEs rapidly decreases as redshift increases,  due to both
lower star formation rate and smaller H{\sc ii} bubbles.  Our model
predicts future wide deep surveys with next generation telescopes,
such as JWST, E-ELT and TMT, can detect LAEs at $z \sim 10$ with a
number density of $n_{\rm LAE} \sim {\rm a~few~} \times10^{-6} ~\rm
Mpc^{-3}$ for the flux sensitivity of $10^{-18} ~\rm
erg\, cm^{-2}\, s^{-1}$.  
By combining these surveys with future 21-cm observations, it could be possible to detect both
LAEs with $\La \gtrsim 10^{42}~\rm erg~s^{-1}$ and their associated giant
H{\sc ii} bubbles with the size $\gtrsim 250 ~\rm kpc$ at $z \sim 10$.
\end{abstract}

%
%

%
%
\section{Introduction}
One of the major challenges in today's astronomy is revealing cosmic
reionization history with galaxy evolution.  Recent CMB observations
suggested cosmic reionization occurred
at $z \sim 8 - 11$ \citep{Komatsu11, Planck14, Planck16}.
Gunn-Peterson tests by the observations of high-redshift QSOs indicated
 cosmic reionization completed at $z \sim 6$ \citep[e.g.,][]{Fan06a}.
However, the ionization history of integer-galactic medium (IGM) has not
been understood yet.  Recent observations of
high-redshift galaxies at $z \gtrsim 7$ are gradually unveiling the cosmic
star formation history \citep{Ouchi10, Bouwens12, Bouwens15,
Oesch15, Oesch16}, and have allowed us to speculate the
cosmic reionization history \citep{Robertson15}.
Yet, even considering the above observational
constraints, various reionization histories remain viable \citep{Cen03, Yajima15b}.  One
of the main uncertainties is low-mass galaxy formation with the
halo mass $\Mh$ less than $\sim 10^{12}~\Msun$.  Due to their higher
number density and the higher escape fraction of
ionizing photons from them \citep{Razoumov10, Paardekooper13, Yajima11, Yajima14c},
 low-mass galaxies can be responsible for main ionizing
sources. However, the detection sensitivities of current observations
are not sufficient to constrain the formation of
low-mass galaxies.  Therefore it is important to theoretically
investigate the formation of low-mass galaxies and
their contribution to cosmic reionization.

Most of high-redshift low-mass galaxies are observed as $\lya$ emitters
 \citep[LAEs;][]{Hu96, Steidel00, Iye06, Gronwall07, Ouchi08, Ouchi10, Bond11, Ciardullo12, Yamada12a}.
\citet{Ouchi10} indicated that LAEs were hosted in halos with the
halo mass $\Mh \sim 10^{11}~\Msun$ by the clustering analysis \citep[see also,][]{Gawiser07}.
\citet{Verhamme08} suggested that LAEs were not dust-enriched well yet so that $\lya$ photons escaped from galaxies against dust attenuation
\citep[see also,][]{Yajima14c}.
These suggest LAEs are likely to be in the early phase of galaxy evolution \citep[e.g.,][]{Mori06}. 
In addition, LAEs are one of major populations of galaxies contributing to the cosmic star-formation rate density in the early Universe \citep{Ciardullo12}.

LAEs also have been used as a tool to
investigate the early Universe \citep{Iye06, Vanzella11, Ono12, Shibuya12, Finkelstein13, Zitrin15, Song16b}.  
Number density of LAEs rapidly decreases at $z \gtrsim 7$ \citep{Ono12, Konno14},
indicating that $\lya$ fluxes from some galaxies were significantly attenuated due to 
neutral hydrogen in the IGM \citep[e.g.,][]{Kashikawa06}. 
Meanwhile, LAEs themselves could provide sufficient ionizing
photons into the IGM \citep{Yajima09, Yajima14c}, and
 some of them could make giant H{\sc ii} bubbles that allowed $\lya$ photons to reach us. 
In this case, galaxies can be observed as LAEs. 
In practice, the most distant LAE has been observed even at $z=8.68$ \citep{Zitrin15}.
Thus, LAEs can be the key objects in understanding the galaxy formation and cosmic reionization.

In this work, we investigate the evolution of LAEs with their
associated H{\sc ii} bubbles.  By modeling 
both LAEs and H{\sc ii} bubbles simultaneously, we estimate $\lya$ luminosity functions (LFs), number
density of LAEs, size distribution of H{\sc ii} bubbles, and the
relation between $\lya$ flux and the size of H{\sc ii} bubble.  These
estimations are useful for future observational
missions.  James Webb Space Telescope (JWST) is going to be launched at
2018, and aims to observe galaxies at $z \sim 10$. Its high
sensitivity of spectroscopy will make it possible to
detect $\lya$ flux from galaxies if they distribute in giant H{\sc ii}
bubbles. Later on, 30-m class ground telescopes, the European Extremely Large Telescope (E-ELT),
the Thirty Meter Telescope (TMT), and the Giant Magellan Telescope (GMT), will investigate galaxies at $z \sim 10$ statistically.  In
addition, several 21-cm observational missions are on going, e.g.,
the LOw Frequency ARray \citep[LOFAR; ][]{Harker10}, and Murchison Widefield Array  \citep[MWA; ][]{Lonsdale09}.  In future, Square Kilometer Array phase-2 \citep[SKA-2;][]{Dewdney09} will
perform large-scale 21-cm tomography.  Since 21-cm
emission comes from H{\sc i} gas, detections of holes of 21-cm signal
indicate giant H{\sc ii} bubbles around galaxies.  Therefore, a
combination of 21-cm and galaxy observations will provide fruitful
information about the cosmic reionization and galaxy formation \citep[e.g.,][]{Lidz09}.

Theoretically large-scale simulations showed that galaxies made
patchy ionization structures in an
inside-out fashion \citep[e.g.,][]{Iliev06a, Iliev12, Mellema06b,Trac07, Ocvirk16}, 
whereas low-density void
regions would be ionized first, in the so-called
outside-in fashion, if AGNs were the main ionizing sources.
However, AGNs are unlikely to be main ionizing sources because of the
rapid decrease of observed number density at redshift $z > 4$
\citep[e.g.,][]{Richards06}, although the contribution of faint AGNs is still under the debate \citep{Madau15, Yoshiura16, Khaire16}.  Thus, in this paper, we model LAEs assuming
the reionization model with the inside-out fashion. 
 In order to understand cosmic reionization theoretically, 
we need high spatial resolution to follow the detailed physical processes of interstellar medium,
as well as large volume to consider large-scale inhomogeneity of ionization structure.
Even current numerical simulations cannot resolve such
a wide dynamic range. For that reason, in this
work, we semi-analytically investigate the cosmic reionization and
galaxy formation based on simple structure formation models and
radiative transfer calculations.

The distributions of LAEs at $z > 6$ can provide us valuable information about cosmic reionization and galaxy formation \citep[e.g., see the review by][]{Dijkstra14}.
\citet{Furlanetto04} investigated the IGM damping wing absorption of  $\lya$ flux from star-forming galaxies, 
and suggested that the $\lya$ line could be a tool to investigate neutral fraction of IGM. 
\citet{McQuinn07b} calculated the large-scale inhomogeneous ionization structure of IGM by combining $N$-body simulations with post-processing radiation transfer calculations. 
They calculated the spatial distribution of LAEs after IGM transmission using simple line profiles of gaussian or single frequency with velocity offset, and showed that the neutral IGM significantly changed the distribution of LAEs \citep[see also,][]{Mesinger15}.  
Numerical modeling of large-scale IGM ionization structure always suffers from the expensive calculation cost of radiation transfer. 
Therefore, \citet{Mesinger08} developed a semi-numerical method to calculate the ionization structure of IGM, and investigated reionization structures with the volume of $(250~\rm Mpc)^{3}$ in comoving unit.
This semi-numerical method significantly reduces the calculation amount, but simulates the ionization structure accurately \citep{Zahn11}.

These previous works investigated the impacts of neutral IGM on the distributions of LAEs with detailed calculations of the IGM ionization structure. 
In this work, we focus on the following two points: {\it 
(1) What are physical properties of observable LAEs at $z \gtrsim 7$?,
(2) Can LAEs be observed even at $z \gtrsim 10$?}
We investigate these points statistically based on a large galaxy sample and $\lya$ line profile models with a wide range of galactic inflow/outflow velocities, 
which have not been understood well in previous studies.  
The IGM transmission sensitively depends on the line profiles that are related with physical properties of gas in galaxies, e.g., velocity field, H{\sc i} column density \citep{Dijkstra06, Verhamme06, Laursen09a, Yajima12b}. 
Some previous works showed the impacts of neutral IGM on the line profiles. 
Using the semi-numerical method, \citet{Dijkstra11} calculated the IGM transmission to $\lya$ flux from a galaxy with $\Mh \sim 10^{10}~\Msun$ at $z=8.6$, and suggested such a galaxy could be observed even when IGM was not fully ionized. They assumed that the galaxy had an expanding gas shell with the outflow velocity of $200~\rm km~s^{-1}$.
\citet{Mesinger15} investigated the redshift evolution of LAEs fraction in Lyman break galaxies (LBGs)
by combining the large-scale ionization structures calculated by the semi-analytical models
and simple Gaussian line profiles with the velocity offset $\Delta v=0, 200$ and $400~\rm km~s^{-1}$.
\citet{Choudhury15} also used a Gaussian profile with a velocity offset depending on the redshift, $\Delta v=100[(1+z)/7]^{-3}~\rm km~s^{-1}$,
and estimated the $\lya$ equivalent width distributions to reproduce the observations. 
Thus most previous works have used simplified models for the line profiles, 
while the ionization structures have calculated in detail by numerical simulations. 
Meanwhile, \citet{Jensen13} estimated outflow velocities from hydrodynamics simulations, 
and claimed the cosmic reionization rapidly completed at redshifts between $z \sim 6$ and $\sim 7$. 
However, it is generally difficult to accurately model galactic outflow  even with state-of-the-art hydrodynamics simulations because the limited numerical resolution does not allow us to determine the strength of supernova feedback. 
Therefore, as a complementary work to previous works, 
here we investigate the physical properties of LAEs, in particular, the typical outflow/inflow velocity
by using wide-range velocity parameter sets from $-300$ to $300~\rm km~s^{-1}$. 
Moreover, in simulating intrinsic $\lya$ line profiles, 
we utilize two spherical symmetric gas models: (1) expanding uniform clouds, (2) expanding shells. 

In addition, we make relations among $\lya$ luminosities, sizes of H{\sc ii} bubbles, 
star formation rates, and stellar masses. 
Recent observations have detected some distant galaxy candidates at $z \gtrsim 8$ 
by the LBG selection \citep[e.g.,][]{Oesch15, Oesch16}. 
They showed the stellar mass and star formation rates of the candidates by SED fittings. 
Therefore, the relations between $\lya$ luminosity and the physical properties
can be a powerful tool to find out the objects which are likely to emit the $\lya$ flux
for future deep spectroscopy observations.  

Then, using modeled LAEs, we also study observability of LAEs at $z \gtrsim 10$ by next generation telescope, e.g., JWST. 
The detection of emission lines from distant galaxies is quite important to determine the redshift 
and confirm the authenticity of candidates.  
Recent observations have allowed us to detect a distant LAE up to $z=8.68$ \citep{Zitrin15}.
However, as shown in recent LAEs observations \citep[e.g.,][]{Ono12},
number density of LAEs rapidly decreases toward higher redshifts at $z \gtrsim 7$. 
This implies the difficulty of observations of LAEs at $z \gtrsim 7$ due to the IGM attenuation when the neutral fraction of the Universe is high. 
Therefore, it is important to study whether we can observe LAEs even at $z \gtrsim 10$ by 
JWST and other upcoming telescopes. 
We will show number densities of observable LAEs at $z \gtrsim 10$ based on simple expanding gas shell/cloud models. 

In this work, we model star formation histories of galaxies based on halo merger histories, 
and ionized bubbles as expanding isolated ones associated with individual galaxies. 
Therefore, we do not consider expansion of ionized bubbles due to the overlap of them, 
i.e., we focus on field galaxies, not clustering galaxies that are likely to make the overlapped giant H{\sc ii} bubbles. 
\citet{McQuinn07a} presented cosmological simulations of different reionization scenarios with post-processing radiation transfer simulations. 
Depending on physical parameters, i.e., star formation efficiency and escape fraction of ionizing photons, various ionization structure and history can be considered under the same value of volume-weighted mean ionization fraction.
They showed the overlaps of H{\sc ii} regions were delayed in the model that assumed low-mass galaxies contributed to cosmic reionization significantly.
In addition, recent CMB observation obtained the Thomson scattering optical depth of IGM $\tau_{\rm e} = 0.058$,
corresponding to the reionization redshift of $z_{\rm re} \sim 8.8$ under the assumption that
all IGM were instantaneously ionized at $z_{\rm re}$. 
Since the Gunn-Peterson tests using QSOs indicated that the reionization was completed at $z \sim 6$ \citep[e.g.,][]{Fan06a}, H{\sc ii} bubbles might not be overlapped well at $z \gtrsim 7$. 
Thus, in this work, we simply follow evolutions of isolated expanding H{\sc ii} bubbles without the effect of overlap. With a simple prescription, we also discuss the effect of the overlap of H{\sc ii} bubbles on the IGM transmission. 

\citet{Tilvi11} modeled LAEs at $z=3-7$ based on halo merger histories from cosmological $N$-body simulations. They assumed that $\lya$ luminosity of each galaxy was simply proportional to SFR and all $\lya$ photons escaped from galaxies, and all ionizing photons were absorbed by interstellar gas. 
On the other hand, in this work, we focus on LAEs at $z > 6$, and consider non-zero escape fraction of ionizing photons to reproduce the observed Thomson scattering optical depth of IGM, 
and IGM transmission depending on $\lya$ line profiles.

We use the cosmological parameters, $\Omega_{\Lambda}=0.7$, $\Omega_{\rm M}=0.3$, $\Omega_{\rm b}=0.045$ and $h=0.7$\citep{Komatsu11, Planck16}.

The paper is organized as follows. We describe our models of star
formation and H{\sc ii} bubbles in \S2.  In \S3, we present the results,
which include the $\lya$ properties, LFs of LAEs and
redshift evolution of number density of LAEs.  We discuss the relation
between the $\lya$ flux of galaxies and the sizes of H{\sc ii} bubbles in
\S4, and summarize in \S5.

%
%
\section{Model}
\label{sec:model}

In the current standard picture of structure formation, halos grow via minor/major mergers. 
In our model, the star formation rate (SFR) in a halo is assumed to be proportional to the growth rate of the halo as follows:
\begin{equation}
{\rm SFR} = \frac{d\Mstar}{dt} = \frac{d\Mstar}{d\Mh}\frac{d\Mh}{dt}.
\end{equation}
The relation between stellar and halo mass can be written as 
$\frac{d\Mstar}{d\Mh} = \frac{d{\rm log}\Mstar}{d{\rm log}\Mh} \frac{\Mstar}{\Mh}$.
The abundance matching analysis by \citet{Behroozi13} indicated that
$\frac{d{\rm log}\Mstar}{d{\rm log}\Mh} \sim {\rm const}$ for $\Mh \lesssim
10^{12}~\Msun$.  
In addition, \citet{Behroozi15} showed the weak dependence of $\frac{\Mstar}{\Mh}$ on halo mass and redshift at the mass range $\Mh \sim 10^{11} - 10^{13}~\Msun$, although $\frac{\Mstar}{\Mh}$ increases with $\Mh$ at $\Mh \lesssim 10^{11}~\Msun$ \citep[see also,][]{Moster13, Kravtsov14}. 
Therefore we here assume $\frac{d\Mstar}{d\Mh} \sim {\rm const}$.
Thus, we estimate SFR from
the growth rate of halo mass with a constant tuning parameter $\alpha$,
i.e., ${\rm SFR} = \alpha \frac{d\Mh}{dt}$.  

\begin{figure}
\begin{center}
\includegraphics[scale=0.4]{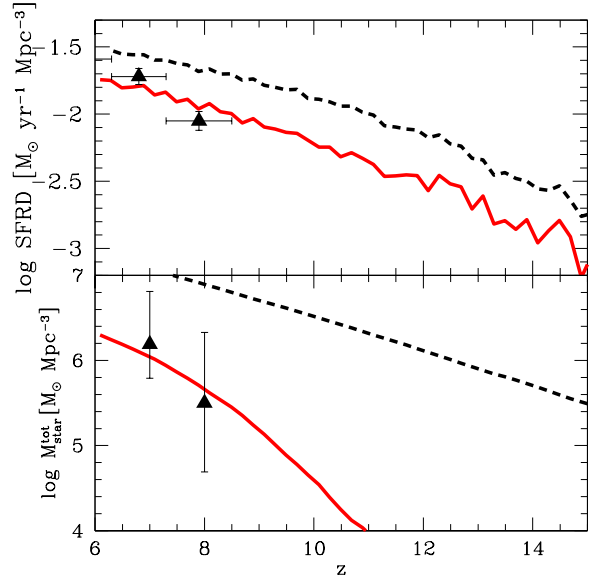}
\caption{
{Upper panel:} Star formation rate density. Red line represents our modeled star formation history based on the halo merger trees. 
Triangle symbols show the observation by \citet{Bouwens15}. The observed star formation rate densities are estimated by integrating 
the luminosity functions in the range of $M_{\rm UV} \le -17$. Our model (red line) also consider only the galaxies brighter than the same limiting magnitude.
{Lower panel:} Stellar mass density. Red line is the cumulated stellar mass of our model considering only the galaxies with $\Mstar \ge 10^{8}~\Msun$. 
Triangle symbols show the observation by \citet{Song16a}, who integrated the derived stellar mass functions for galaxies with $\Mstar \ge 10^{8}~\Msun$.
Black dash lines represent the star formation rate and stellar mass densities considering all galaxies without the thresholds equivalent to the observations.
}
\label{fig:sfr}
\end{center}
\end{figure}

In order to estimate the
growth of halos, we use halo merger trees based on an extended
Press-Schechter formalism \citep{Somerville99, Khochfar01, Khochfar06}.  The halo merger trees include 5000 realizations
with the halo mass range of $10^{9} - 10^{13}~\Msun$ at $z=6$.  In this
work, we allow star formation only for halos with the mass $\Mh >
10^{8}~\Msun$, because star formation in less
massive halos can be significantly suppressed due to UV background or
internal stellar feedback \citep[e.g.,][]{Okamoto08a, Hasegawa13}.  In deriving statistical properties, e.g., cosmic SFR density,
stellar mass density and LF, we sum the
contribution of each merger tree with normalization factors
that reproduce the halo mass function of \citet{Sheth02}.  We determine the tuning parameter by
using observed cosmic SFR density and stellar mass
density at $z \sim 7 - 8$.
Figure~\ref{fig:sfr} shows our modeled SFR and stellar mass density with
observations.  We choose the parameter $\alpha$ by the least square
fitting to the four points of the observations.  The observed SFR
and stellar mass densities consider only galaxies with $M_{\rm UV}
< -17$  and
$\Mstar \ge 10^{8}~\Msun$, respectively. Therefore,
we consider only the galaxies satisfying the above criteria for the fitting.  
The best fit value is $\alpha=3.3 \times
10^{-3}$.  Note that, so far there is no available data
about $\frac{d\Mstar}{d\Mh}$ at $z \gtrsim 9$ and $\alpha$ can change
with halo mass and redshift.  However, for simplicity, we assume
$\alpha$ is constant. Following the above way, we derive
star formation history from each halo merger tree.
Although we consider only bright galaxies in deriving the parameter $\alpha$ via the comparisons with the observations, 
we follow star formation of all halos with $\Mh \ge 10^{8}~\Msun$ in studying cosmic reionization and LAEs in this work.
Black dash lines in the panels of Figure~\ref{fig:sfr} represent the total SFR and stellar mass densities as a function of redshift.  
The top panel of Figure~\ref{fig:sfr} shows that low-mass faint galaxies have non-negligible contribution to the cosmic SFR density and are expected to contribute to cosmic reionization as recent studies indicated
\citep[e.g.,][]{Yajima09, Yajima11, Yajima14c, Robertson13, Cai14, Wise14}.

\begin{figure}
\begin{center}
\includegraphics[scale=0.4]{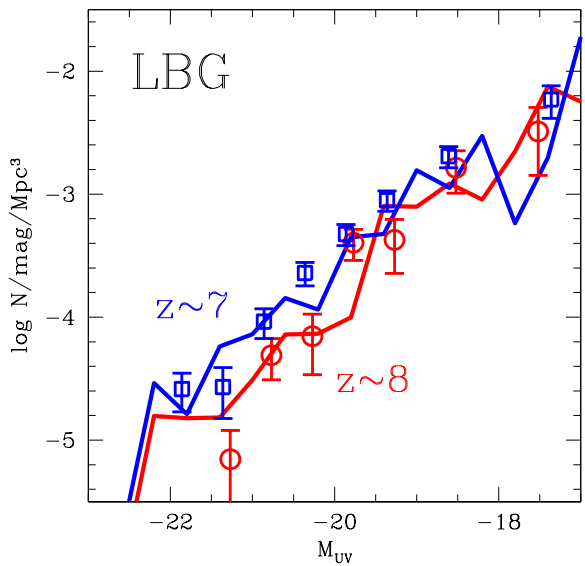}
\caption{
UV luminosity functions. Open squares and circles represent observed LBGs at $z\sim7$ and $\sim 8$ by \citet{Bouwens15}.
}
\label{fig:LBG}
\end{center}
\end{figure}

We also derive UV LFs by converting SFR
to UV flux with $L_{\rm \nu, UV} = 0.7 \times 10^{28}~{\rm
erg~s^{-1}}~ \left( \frac{\rm SFR}{\rm 1~\Msunyr} \right) $ \citep{Madau99}.  Figure~\ref{fig:LBG} shows our modeled LFs at
$z \sim 7$ and $\sim 8$ with those from the recent observation by \citet{Bouwens15}.  The
observation indicated that a part of UV flux are absorbed by dust
with the escape fraction $\fescuv \sim 0.6$.  Our
modeled LFs match the observation well with the same UV escape fraction.

\begin{figure}
\begin{center}
\includegraphics[scale=0.4]{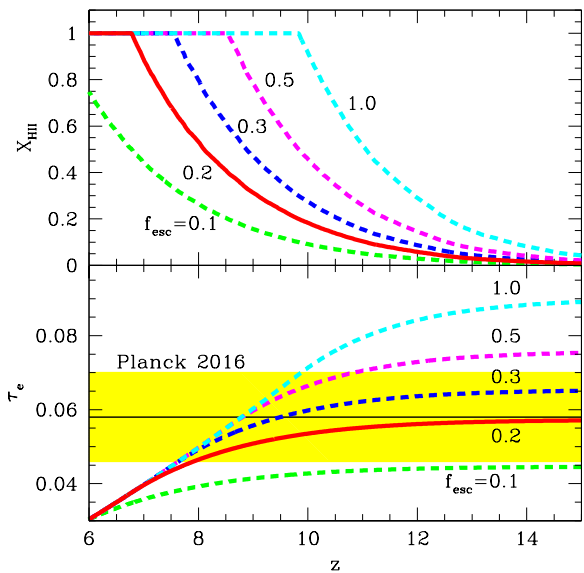}
\caption{
{ Upper panel:} Ionization history. Different lines show ionization histories with different escape fractions of ionizing photons. 
{ Lower panel:} Thomson scattering optical depth.  Different lines are estimated with different ionization histories presented in the upper panel. 
Yellow shade represents the estimation by the CMB observation \citep{Planck16}. 
}
\label{fig:tau}
\end{center}
\end{figure}

Galaxies ionize the IGM as star formation proceeds. 
In a one-zone approximation, we estimate time evolution of cosmic ionization degree \citep{Barkana01}, 
\begin{equation}
\frac{d\Qhii}{dt} =  \frac{1}{n_{\rm H}^{0}} \nion \fescion - \alpha_{\rm B} C (1+z)^{3} n_{\rm H}^{0} \Qhii,
\end{equation}
 where $\Qhii$ is the volume fraction of H{\sc ii},
$n_{\rm H}^{0}$ is the present-day hydrogen number density ($\sim 2
\times 10^{-7}~\rm cm^{-3}$), 
$\nion$ is the intrinsic ionizing photon emissivity per unit volume, 
$\alpha_{\rm B}$ is the case-B recombination rate, 
$C$ is a clumpiness factor of IGM, and  $\fescion$ is escape fraction of
ionizing photons, which is a free parameter here.  We
estimate $\nion$ from star formation history of each halo by using a population synthesis code
{\sc starburst99} \citep{Leitherer99} with the Salpeter initial mass function with the metallicity
of $10^{-2}~\rm Z_{\odot}$. The reionization history is shown in the
upper panel of Figure~\ref{fig:tau}.  As $\fescion$ increases, the IGM is
ionized earlier.  We assume the recombination rate for $T=10^{4}~\rm K$
($\alpha_{\rm B} = 2.6 \times10^{-13}~\rm cm^{3}\;s^{-1}$) and
$C=3$, as suggested by numerical simulations
\citep{Pawlik09b, Jeon14}.

The cosmic reionization history is regulated by
$\fescion$.  Free electrons produced by the cosmic reionization
 contribute to the Thomson scattering optical depth
($\tau_{\rm e}$) of CMB photons, defined as
\begin{equation}
\tau_{\rm e} = \int_{0}^{z_{\rm rec}} \sigma_{\rm T} n_{\rm e}(z)  c \left| \frac{dt}{dz} \right| dz,
\end{equation}
where $z_{\rm rec}=1100$ is the redshift at the time of recombination. 
%
%
In this work, we assume the single ionization fraction of helium is the same
as the one for hydrogen at $z \ge 3$, and the double ionization takes
place at $z < 3$ \citep[e.g.,][]{Wyithe10, Inoue13}.  Recent simulations
show that indeed the fraction of He{\sc ii} is close to the H{\sc ii}
fraction at high redshifts, although the ionization fraction of
helium is slightly lower than the one for hydrogen \citep{Ciardi12}.
The top panel of Figure~\ref{fig:tau} represents the ionization history
of hydrogen gas with the different $\fescion$.  The bottom panel
of Figure~\ref{fig:tau} shows $\tau_{\rm e}$ with different $\fescion$.
The Thomson scattering optical depth $\tau_{\rm e}$ increases with
redshift in a way depending on
$\fescion$ due to the different ionization histories.  We find
$\fescion=0.2$ nicely reproduces the CMB observation
(Planck 2016).  Therefore, in this work, we adopt
$\fescion=0.2$ with no redshift evolution.  In fact, \citet{Yajima14c} showed $\fescion$ is constant with redshift and
$\sim 0.2$  by cosmological simulations
with radiative transfer calculations.  

%
%

\section{Results}
\label{sec:result}

\subsection{Evolution of ionized bubbles around galaxies}

\begin{figure}
\begin{center}
\includegraphics[scale=0.43]{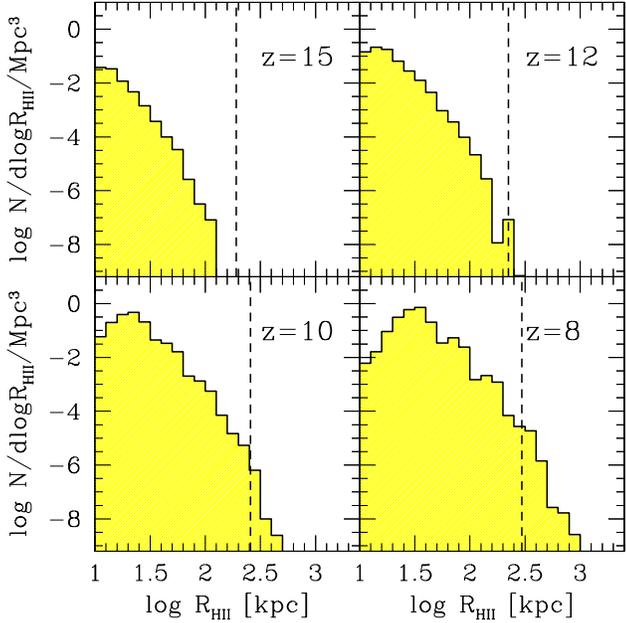}
\caption{ Probability distribution functions of sizes of H{\sc ii} bubbles
associated with individual galaxies.  Vertical dash lines
are corresponding to the viewing angle of 1 arcmin.}
\label{fig:rhii_hist}
\end{center}
\end{figure}

We follow the growth history of H{\sc ii} bubbles around individual galaxies with 
the star formation efficiency $\alpha$ and $\fescion$ estimated above. 
Sizes of H{\sc ii} bubbles evolve with ionizing photon emissions, recombination, and cosmic expansion
as follows \citep{Cen00}:
\begin{equation}
\frac{d\rhii^{3}}{dt} = 3H(z)\rhii^{3} + \frac{3\Nion \fescion}{4 \pi \nh(z)} - C\nh(z)\alpha_{\rm B} \rhii^{3},
\end{equation}
where $H(z)$ is the Hubble constant at specific redshifts, and $\Nion$
is intrinsic ionizng photon emissivity of each galaxy.  The ionizing fronts can
propagate up to the Str$\rm \ddot{o}$mgren radius during the recombination time scale
\citep{Spitzer78}.  
The recombination time scale is $t_{\rm rec} \sim
\frac{1}{\alpha_{\rm B} \nh} \sim 0.5 ~{\rm Gyr} ~\left( \frac{1+z}{11}
\right)^{-3}$, which is longer than the typical time scale
 over which SFR changes more than factor $\sim 2$.
Therefore, H{\sc ii} bubbles do not reach the equilibrium state,
and we need to consider SFR history to estimate the sizes of H{\sc
ii} bubbles at given redshifts.  We estimate
probability distribution function (PDF) of the sizes of ionized bubbles
($\rhii$) as shown in Figure~\ref{fig:rhii_hist}.  In our model, higher
mass halos tend to possess larger ionized bubbles.  Due to
the decrease in the number density of halos
on the high-mass end of a halo mass function \citep{Sheth02},
the PDF of $\rhii$ rapidly decreases at larger $\rhii$.  Although the
ionizing front does not reach the size of Str$\rm \ddot{o}$mgren sphere
$r_{\rm st} \propto (\Nion)^{1/3} (1+z)^{-2}$, it can be used as a
rough indicator of sizes of H{\sc ii} bubbles.  As redshift
decreases, the IGM density decreases while the number density of
massive halos with higher ionizing photon
emissivity increases. 
Therefore, the tail of PDF in the
large-$\rhii$ end shifts to larger $\rhii$ at lower redshift.  Future
21 cm observations, e.g., SKA-2, is supposed to
probe the IGM ionization structure with the angular resolution of $\sim 1'$.
Therefore, at $z \lesssim 10$, the tail of
PDF at large $\rhii$ can be observationally
investigated in future.
The halo number density monotonically increases as the halo mass
decreases at a fixed redshift.  Since the size of H{\sc
ii} bubble is positively related with halo mass as will be shown in Section
3.3, it seems that the PDFs in the small-$\rhii$ end monotonically
increase as $\rhii$ decreases.  However, there are peaks in the
PDFs, below which they decrease as $\rhii$ decreases.  This is
caused by the threshold of halo mass for star formation
imposed in this work.

Note that, in this work, we do not take the effect of overlap of H{\sc ii} bubbles into account. 
The galaxy clustering causes the overlap of H{\sc ii} bubbles that can extend the tails of PDFs to larger size. 
Since the overlapping effect can be larger as redshift decreases, 
the slopes of PDFs at lower redshifts can be changed to be more shallower.  
We will discuss the impacts of the overlap of H{\sc ii} bubbles on luminosity functions in Section~\ref{sec:LFlae}. 

\subsection{$\lya$ luminosity functions}
\label{sec:LFlae}

Absorption of ionizing photons by interstellar medium within
galaxies results in $\lya$ emissions via recombination processes, while
escaped photons cause the cosmic reionization as shown in the previous
section.  The $\lya$ luminosity ($\La$) is estimated by
\begin{equation}
\La = 0.68 (1.0 - \fescion) \fesca \epsilon_{\lya} \Nion, 
\end{equation}
where $\fesca$ is the escape fraction of $\lya$ photons from galaxies,
$\epsilon_{\lya}=10.2~\rm eV$ is the energy of a $\lya$ photon.  The
escape fraction of $\lya$ photons $\fesca$ can be lower than $\fescuv$
because the path length of $\lya$ photons until escape can be longer due
to multiple scattering process.
However, if dust mainly distributes in H{\sc
i} gas clumps, $\fesca$ does not become lower than $\fescuv$ because
$\lya$ photons are scattered by hydrogen on surface of the clumps before
interacting with dust.  \citet{Ciardullo12} indicated that $\fesca
\sim \fescuv$ for observed LAEs at $z \sim 2$.  In addition,
cosmological simulations of \citet{Yajima14c} showed that $\fesca$
was $\gtrsim 0.6$ and similar to $\fescuv$ at
$z \gtrsim 6$.  Therefore, in this work, we assume that
$\fesca=\fescuv=0.6$.

Next, we estimate IGM transmission as a function of
wavelength.  As in \citet{Cen00}, we divide the
paths along which $\lya$ photons
travel from galaxies to us into the two parts, i.e.,
outside and inside ionized bubbles, and separately estimate each contribution.
The transmission outside ionized bubbles is estimated as follows:
\begin{equation}
\tau(\lambda_{\rm obs}, z_{\rm s}) = \int_{z_{\rm r}}^{z_{\rm i}}  
dz\, c\, \frac{dt}{dz}\, \nh(z) x_{\rm HI}
\sigma_{\lya}\, [\lambda_{\rm obs} / (1 + z)],
\label{eq:tau}
\end{equation}
where
$z_{\rm i} \sim z_{\rm s} - \frac{\rhii \times (1+z)}{R_{\rm H}}$.
%
Here, $z_{\rm r}$ is redshift when the cosmic reionization
completes, which we set $z_{\rm r}=6$,
$z_{\rm i}$ is redshift when $\lya$ photons
pass through ionizing front, $z_{\rm s}$ is redshift of galaxy,
$\sigma_{\lya}$ is the scattering cross section for H{\sc i} gas,
and $R_{\rm H}$ is the size of the cosmological horizon at $z_{\rm s}$. 
We assume the outside of the bubble is completely neutral, i.e., $x_{\rm HI}=1$.
The $\sigma_{\lya}$ is estimated by \citep{Verhamme06}
\begin{equation}
\sigma_{\lya}[\lambda] = 1.041 \times 10^{-13} \left( \frac{T}{10^{4} ~\rm K}\right)^{-\frac{1}{2}} 
\frac{H(x,a)}{\sqrt{\pi}}.
\end{equation}
We set $T=10^{4}~\rm K$ in this work. 
Here, $H(x,a)$ is the Voigt function,
\begin{equation}
H(x,a) = \frac{a}{\pi} \int_{-\infty}^{+\infty} \frac{e^{-y^{2}}}{(x-y)^{2} + a^{2}} dy
\end{equation}
where 
$x \equiv (\nu - \nu_{0})/\Delta \nu_{\rm D}$,
$\nu_{0}=2.466 \times 10^{15}~\rm Hz$ is the line-center frequency,
$\nu_{\rm D}$ is the Doppler width,
$a=\Delta \nu_{\rm L}/(2 \Delta \nu_{\rm D})$,
$\Delta \nu_{\rm L}=9.936\times10^{7}~\rm Hz$ is the natural line width. 
We here use the fitting formula of $H(x,a)$ given by \citet{Tasitsiomi06}. 

Even inside ionized bubbles, a tiny fraction of neutral hydrogens exist.  We estimate the IGM transmission from
ionizing front to virial radius 
with the neutral fraction under the ionization equilibrium state,
\begin{equation}
x_{\rm HI} = 1.5 \times 10^{-5} \left( \frac{C}{3} \right)
\left( \frac{r}{\rm kpc} \right)^{2}
\left( \frac{\Nion}{10^{50}~\rm s^{-1}} \right)^{-1}
\left( \frac{1+z}{8} \right)^{3}.
\end{equation}
Note that, the optical depth outside the ionized bubble is dominant in our work. 
The IGM transmission is mostly almost zero at $\lambda \lesssim \lambda_{0}$, 
where $\lambda_{0}=1216~\rm \AA$ is the wavelength of $\lya$ line center.

Considering the IGM transmission, we derive $\lya$ LFs, and compare
them with the observation of \citet{Konno14}.  Depending on the
shape of $\lya$ line profile, the IGM transmission significantly
changes.  Even with recent deep spectroscopies, however, it is
difficult to determine intrinsic $\lya$ line profiles, i.e., before the
IGM extinction \citep[e.g.,][]{Ouchi10}.  
The intrinsic $\lya$ line profile depends on the physical nature of galaxies, e.g., H{\sc i} column density and 
velocity field \citep[e.g.,][]{Verhamme06}. 
In this work, we calculate intrinsic $\lya$ line profiles by $\lya$ radiation transfer simulations using the code developed in \citet{Yajima12b}, and study the physical nature of LAEs through the comparison with the observation \citep{Konno14}.
For the purpose, we employ the following two models for the internal velocity structure of a galaxy: (1) expanding cloud, (2) expanding shell. Both cloud and shell models have only two parameters: (1) H{\sc i} column density $N_{\rm HI}$, (2) outflow velocity $V_{\rm out}$. 
The intrinsic line profiles of the cloud model are calculated based on spherically outflowing gas with the following velocity structure,
\begin{equation}
V(r) = V_{\rm out} \left( \frac{r}{R_{\rm edge}} \right),
\end{equation}
where $R_{\rm edge}$ is the edge of the spherical cloud and $V_{\rm out}$ is the outflow velocity at $R_{\rm edge}$. Here uniform gas density in the cloud is assumed.
Although the simulated profiles do not depend on $R_{\rm edge}$, the source size is likely to increase with $R_{\rm edge}$.
Since the surface brightness decreases with the source size, 
the detectability of $\lya$ flux depends on the choice of $R_{\rm edge}$. 
Nevertheless, in this work, we assume that LAEs are compact and no flux is lost
based on theoretical motivation: 
the typical sizes of galaxies become $\sim \lambda_{\rm spin} R_{\rm vir}$ \citep[e.g.,][]{Mo98} where $\lambda_{\rm spin}$ is a halo spin parameter and expected to be less than $\sim 0.1$ \citep[e.g.,][]{Bullock01}.
Therefore we consider $R_{\rm edge} = 0.1 R_{\rm vir}$ in which the source size may not be subject to the flux loss.
The uncertainty of the surface brightness will be discussed in Sec.~\ref{sec:modellimit}.

In the shell model, the velocity field is monochromatic, i.e., 
\begin{equation}
V(r)={\rm const.}=V_{\rm out}.
\end{equation}
Our radiation transfer calculations are carried out in 100 spherical shells. 
All shells have same H{\sc i} density in the cloud model, while only most outside shell has H{\sc i} gas in the shell model. 

The cloud and shell models approximate galactic outflows due to stellar feedback. 
When most of gas are being evacuated due to starburst, 
H{\sc i} gas structure may be close to the shell model. 
On the other hand, in the case of smooth star formation history, 
stars keep being formed in static H{\sc i} gas at near galactic centers, while a part of gas are evacuated from galaxies.
In this case, the gas structure may be closer to the cloud model. 
The star formation and gas structure in high-redshift galaxies are still under the debate \citep[e.g.,][]{Kimm14, Hopkins14, Yajima17}. 
Therefore we here study LAEs by using these two models. 

\begin{figure*}
\begin{center}
\includegraphics[scale=0.7]{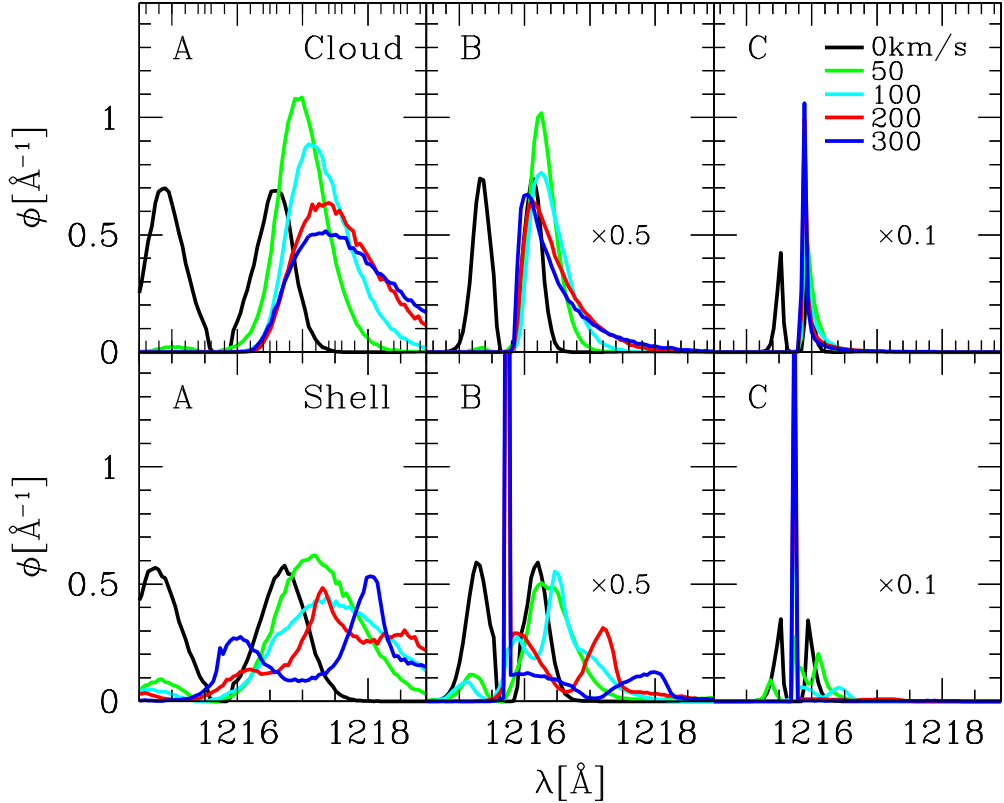}
\caption{
$\lya$ line profiles from expanding spherical gas clouds or shells. 
Upper panels represent the line profiles of the cloud model that assume
the velocity field as
$V(r) = V_{\rm out} \left( \frac{r}{R_{\rm edge}}\right)$, where $R_{\rm edge}$ is 
the radius of clouds and $V_{\rm out}$ is the velocity at $R_{\rm edge}$.
Lower panels present the line profiles of the shell model in which the velocity is constant, 
i.e., $V(r) = V_{\rm out}$.
The different panels from left to right show the $\lya$ profiles considering different 
H{\sc i} column densities: $N_{\rm HI}=2 \times 10^{20}~\rm cm^{-2}$ (panel A),
$2 \times 10^{19}~\rm cm^{-2}$ (panel B), and $2 \times 10^{18}~\rm cm^{-2}$ (panel C). 
The different lines represent the different expanding velocities at $R_{\rm edge}$. 
The $\phi(\lambda)$ is normalized to be unity when it is integrated over the wavelength. 
The $\phi(\lambda)$ of panel (B) and (C) is artificially reduced by a factor 2 and 10. 
}
\label{fig:inprof}
\end{center}
\end{figure*}

Figure~\ref{fig:inprof} shows the modeled line profiles with various outflowing velocities. 
The $\lya$ line profiles of the cloud model get the
asymmetric shape with a peak at redder wavelength for the outflow
velocity field, because $\lya$ photons at bluer wavelength are scattered
by H{\sc i} gas due to the Doppler shift.
As $N_{\rm HI}$ increases, the line profiles are extended, and the peak frequency shift farther from the line center frequency.
In the case of the shell model, the line profiles become more complicated
because of the back scattering effect \citep{Verhamme06}.
Observable photons scattered by the shell at the far side from an observer make a bump at redder wavelength. 
In addition, when the H{\sc i} column density is low and the expanding velocity is large, 
most of $\lya$ photons can directly escape from the shell, resulting in the peaks at the line center frequency
as shown in middle and right panels of the shell model.
For the inflow velocity field, the $\lya$ line profile becomes the mirror
symmetric shape to the one for outflow with the same absolute value of
the velocity.  

\begin{figure*}
\begin{center}
\includegraphics[scale=0.7]{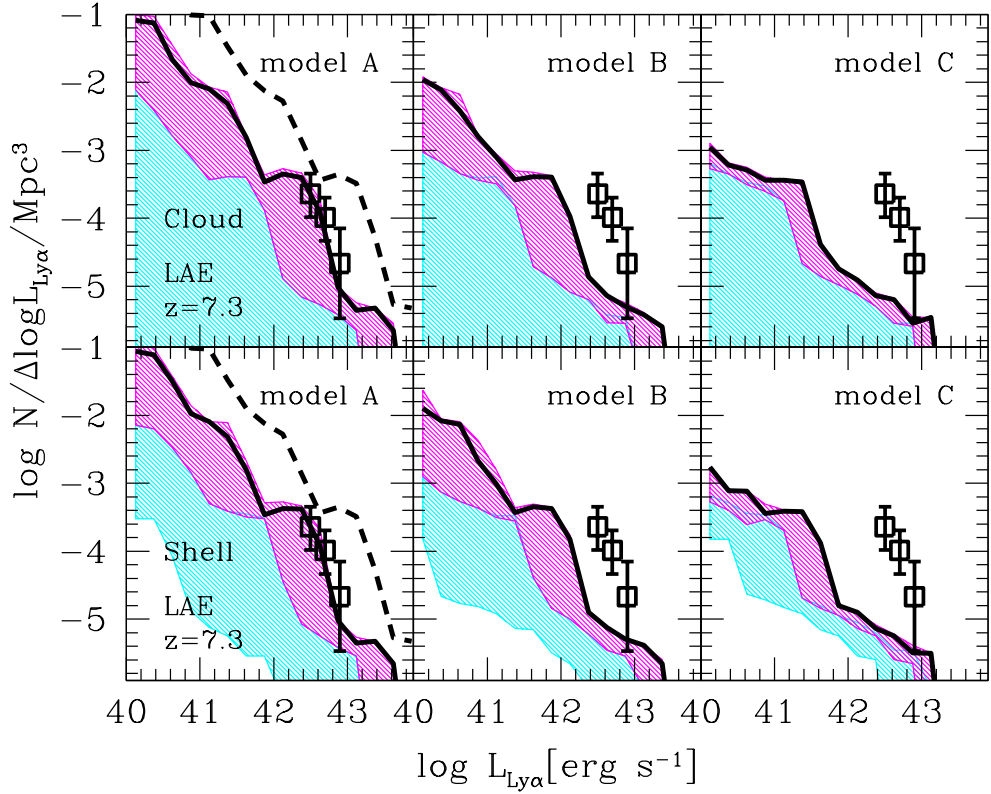}
\caption{
$\lya$ luminosity functions (LFs) at $z=7.3$. 
Square symbols represent the observed LF of LAEs at $z=7.3$ by \citet{Konno14}.
Upper and lower panels present modeled LFs using $\lya$ line profiles of the cloud and shell models, respectively.
Different panels from left to right show modeled LFs based on $\lya$ line profiles to different H{\sc i} column densities. 
Magenta and cyan shades show the range of LFs considering $\lya$ line profile with different outflow and inflow 
velocity with the range $0 \sim \pm 300~\rm km~s^{-1}$.
Black solid lines are best fitted ones to the observation. 
Black dash line in the left panel shows the LF before considering IGM transmission. 
}
\label{fig:LFlae}
\end{center}
\end{figure*}

With the simulated intrinsic line profiles, we can estimate $\lya$
LFs for given $N_{\rm HI}$ and $V_{\rm out}$.  In this work, we infer
typical $N_{\rm HI}$ and $V_{\rm out}$ of LAEs by comparing modeled LFs with the
observed one by \citet{Konno14}.
Figure~\ref{fig:LFlae} show the LFs.  Here we calculate the LFs for
three H{\sc i} column density models, $N_{\rm HI}=2\times10^{20}~\rm cm^{-2}$
(model A), $N_{\rm HI}=2\times10^{19}~\rm cm^{-2}$ (model B), and $N_{\rm
HI}=2\times10^{18}~\rm cm^{-2}$ (model C).
The column density of model A is corresponding to Damped Lyman-$\alpha$ Systems \citep[DLAs:][]{Wolfe05}.
\citet{Yajima12h} showed that
DLAs distributed at lines of sight passing star-forming regions in high-redshift star-forming galaxies
by combining cosmological SPH simulations with radiation transfer calculations.
These column densities are
optically thick to ionizing photons, hence not consistent with
$\fescion=0.2$.  However, recent
simulations showed ionizing photons mostly escape along ionized holes
created by radiative and SNe feedback \citep[e.g.,][]{Yajima09, Yajima11, Kimm14}. Thus, $\fescion$
of 20 $\%$ can be  considered
as the fraction of viewing angle along which star forming regions
are not covered by H{\sc i} gas. 
For simplicity, we do not take account of the effect of such holes on line profiles. 
 Note that, however, $\lya$ line profiles
somewhat change due to the holes, clumpiness or other detailed structure
in H{\sc i} gas \citep{Dijkstra12}.

The shaded regions in Figure~\ref{fig:LFlae} represent the LFs using different
$\lya$ line profiles with the velocity range from $V_{\rm out}= -300~\rm
km~s^{-1}$ to $300~\rm km~s^{-1}$.  
 The best fit velocities for the cloud model are $180~\rm \kms$ (model A), $190~\rm
\kms$ (model B), and $110~\rm \kms$ (model C), 
while in the case of the shell model they are $130~\rm \kms$ (model A), $60~\rm
\kms$ (model B), and $40~\rm \kms$ (model C), 
as summarized in Table~\ref{table:model}.
Only model A with the velocity range $V_{\rm out} \sim 100 - 300~\rm \kms$ can reproduce the observed LF well. 
Therefore we suggest LAEs are likely to have high H{\sc i} column density with $\gtrsim 10^{20}~\rm cm^{-2}$ and outflowing H{\sc i} gas with velocity $\gtrsim 100~\rm km~s^{-1}$.
These outflow velocities are consistent with 
recent observations of [C{\sc ii}] emissions from LAEs at $z \sim 7$ in which they showed the velocity offsets between
$\lya$ and [C{\sc ii}] lines \citep{Pentericci16}.
Note that, the cloud model has low velocity component at a galactic center, unlike the shell model. Therefore, the velocity can not be compared directly between the models.
The $\lya$ profile of the cloud model
monotonically shifts to redder one as the outflow velocity increases in
model A  as shown in Figure~\ref{fig:inprof}.  On the other hand, as the H{\sc i} column density decreases,
the $\lya$ profile moves back to the line center frequency at specific
velocity of $< 300~\rm \kms$.  This is because $\lya$ photons can escape
from the cloud before shifting to longer wavelength due to lower optical depth. 
In this work, the best fit velocities for
model B and C roughly 
correspond to those producing the $\lya$ profiles shifted farthest away.

In the case of the shell model, even for model B and C,  the wavelength of the bump made by the back scattering effect shifts to longer one as the outflow velocity increases.
However, the flux at the line center frequency becomes high as the outflow velocity increases.
Therefore, for lower column density models, $\lya$ flux is efficiently attenuated when the outflow velocity is higher than specific values. 

In addition, the width of line profile becomes
smaller as the H{\sc i} column density decrease.  Even with the best fit
velocities, the number densities of LAEs in the model B and C are smaller than the observation
because of the narrower line profiles resulting in lower IGM
transmission.  Hence, in order to get higher IGM transmission to
reproduce the LF, 
LAEs are likely to have the column density higher than $\sim 10^{20}~\rm cm^{-2}$.
Thus, in this work, we consider model A as our fiducial model.
\citet{Verhamme08} also suggested similar column densities and outflow velocities for LAEs at $z \sim 3$ via comparisons of their modeled $\lya$ line profiles based on the shell model with the observations \citep[but see,][]{Konno16}. 
On the other hand, local LAE analogues are likely to have lower H{\sc i} column densities 
\citep{Henry15, Verhamme15}.

\begin{figure}
\begin{center}
\includegraphics[scale=0.4]{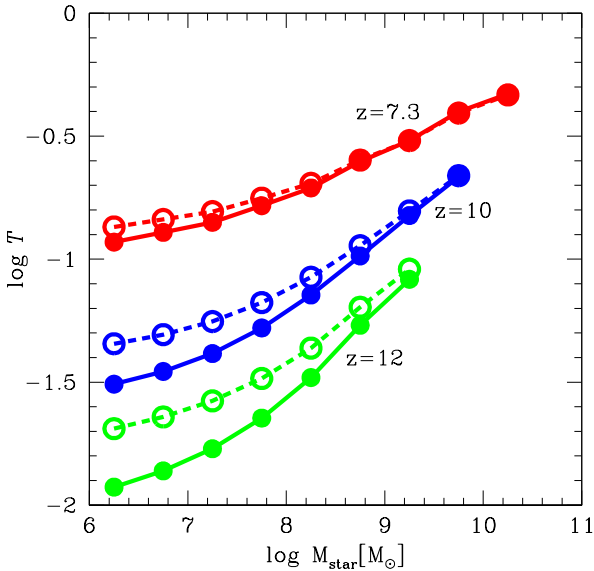}
\caption{
IGM transmission as a function of stellar mass of galaxies.
Filled and open circles represent median values of the cloud and shell models, respectively.
Different colors indicate different redshifts. 
}
\label{fig:trans}
\end{center}
\end{figure}

The IGM transmissions are presented in Figure~\ref{fig:trans}.
Even at $z=7.3$ more than half of intrinsic $\lya$ flux can be attenuated due to residual neutral IGM outside of H{\sc ii} bubbles. 
This is roughly consistent with observations that suggested LAEs fraction in LBGs rapidly decreases from $z \sim 6$ to $z \sim 7$ \citep[e.g.,][]{Ono12, Konno14}. 
\citet{Bolton13} suggested number densities of Lyman-limit systems or DLAs increased with redshift, resulting in the decreased number density of LAEs due to the IGM attenuation \citep[see also,][]{Choudhury15, Mesinger15}.
In addition, recently \citet{Sadoun17} suggested that 
the decreased number density of LAEs can be explained by considering only infalling IGM near virial radius, without the modeling of redshift evolution of the neutral fraction of whole IGM. 
They assumed that the IGM ionization structure was determined by external UVB, 
while the IGM transmission in our models basically results from the neutral IGM outside H{\sc ii} bubbles created by galaxies themselves. 

As discussed in Sec.~\ref{sec:lyamst}, 
the size of H{\sc ii} bubble increases with stellar (or halo) mass. 
Therefore, the transmission increases with stellar mass (see Eq. ~\ref{eq:tau}).
As redshift increases, the typical size of H{\sc ii} bubbles decreases (Figure~\ref{fig:rhii_hist}),
and the mean IGM density increases. 
Thus, the transmission decreases as redshift increases. 
At $z=12$, the transmission of even massive galaxies with $M_{\rm star} \sim 10^{9}~\Msun$ is less than $\sim 10~\%$, 
while even low-mass galaxies have higher transmission than $10~\%$ at $z=7.3$.
In addition, at higher redshifts, the transmission of the shell model becomes higher than the cloud model. 
As shown in Figure~\ref{fig:inprof}, the profiles of the shell model have a bump at longer wavelength due to the back scattering effect.
$\lya$ photons at this bump can penetrate the IGM even when a H{\sc ii} bubble is small at high redshifts.   

\begin{figure}
\begin{center}
\includegraphics[scale=0.4]{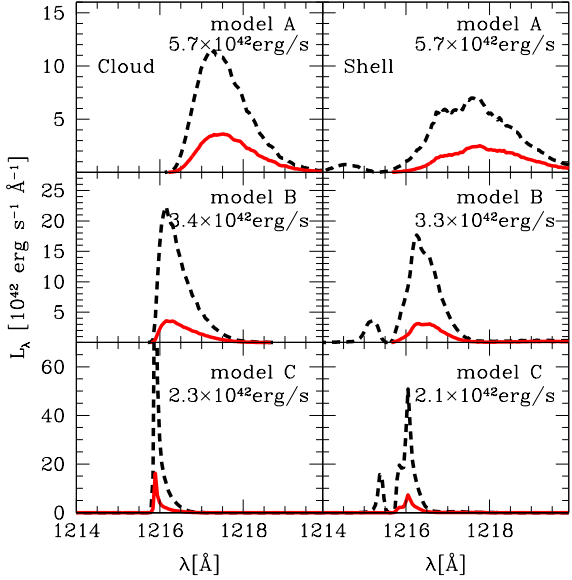}
\caption{
$\lya$ line profiles.
Red solid lines show emergent line profiles of 
a halo of $5.9 \times 10^{11}~\Msun$ at z=7.3,
which evolves to a halo of $1.0\times10^{12}~\Msun$ at $z=6.0$.
Intrinsic $\lya$ luminosity of the halo is $1.3\times10^{43}~\ergs$. 
Left and right panels represent the $\lya$ line profiles of the cloud and shell models.
Emergent luminosities after IGM transmission are shown in the panels.
Black dash lines are the line profiles before considering IGM transmission.
}
\label{fig:prof}
\end{center}
\end{figure}

Emergent $\lya$ line profiles are shown in Figure~\ref{fig:prof}.  As
H{\sc i} column density decreases, intrinsic $\lya$ line profiles become
narrower and peak positions shift to shorter wavelength.  IGM
transmission increases with wavelength because 
photons originally with long wavelength are redshifted and cease to be scattered by the IGM, 
while $\lya$ flux near the line center is reduced
efficiently by the IGM scattering.  FWHMs of the emergent line profiles of the cloud model are $1.5~\rm
\AA$ (model A), $8.9\times10^{-1}~\rm \AA$ (model B) and
$5.2\times10^{-2}~\rm \AA$ (model C).  The emergent line profiles of the shell model have somewhat larger FWHMs.
Therefore it is difficult to
distinguish the different column density models in the current 
spectroscopic observation
with the resolution of $R \sim 1000 - 2000$ \citep[e.g.,][]{Shibuya12}. 
 Future high-dispersion spectroscopies with $R \gg
2000$, e.g., Prime Focus Spectrograph on Subaru or JWST, will be able to reveal the detailed shape of profile.

Line profiles 
in inflowing gas models result in the underproduction of observable LAEs,
because most of $\lya$ photons are scattered by
the IGM. This is consistent with the observation by \citet{Ouchi10},
which indicated LAEs at $z = 6.6$ are likely to have outflowing gas by
the composite spectrum. In addition, \citet{Shibuya14b} measured
outflow velocities of individual LAEs at $z \sim 2$, and indicated that
LAEs were likely to have outflow with $V_{\rm out} \gtrsim 150~\rm
\kms$.

\begin{figure}
\begin{center}
\includegraphics[scale=0.4]{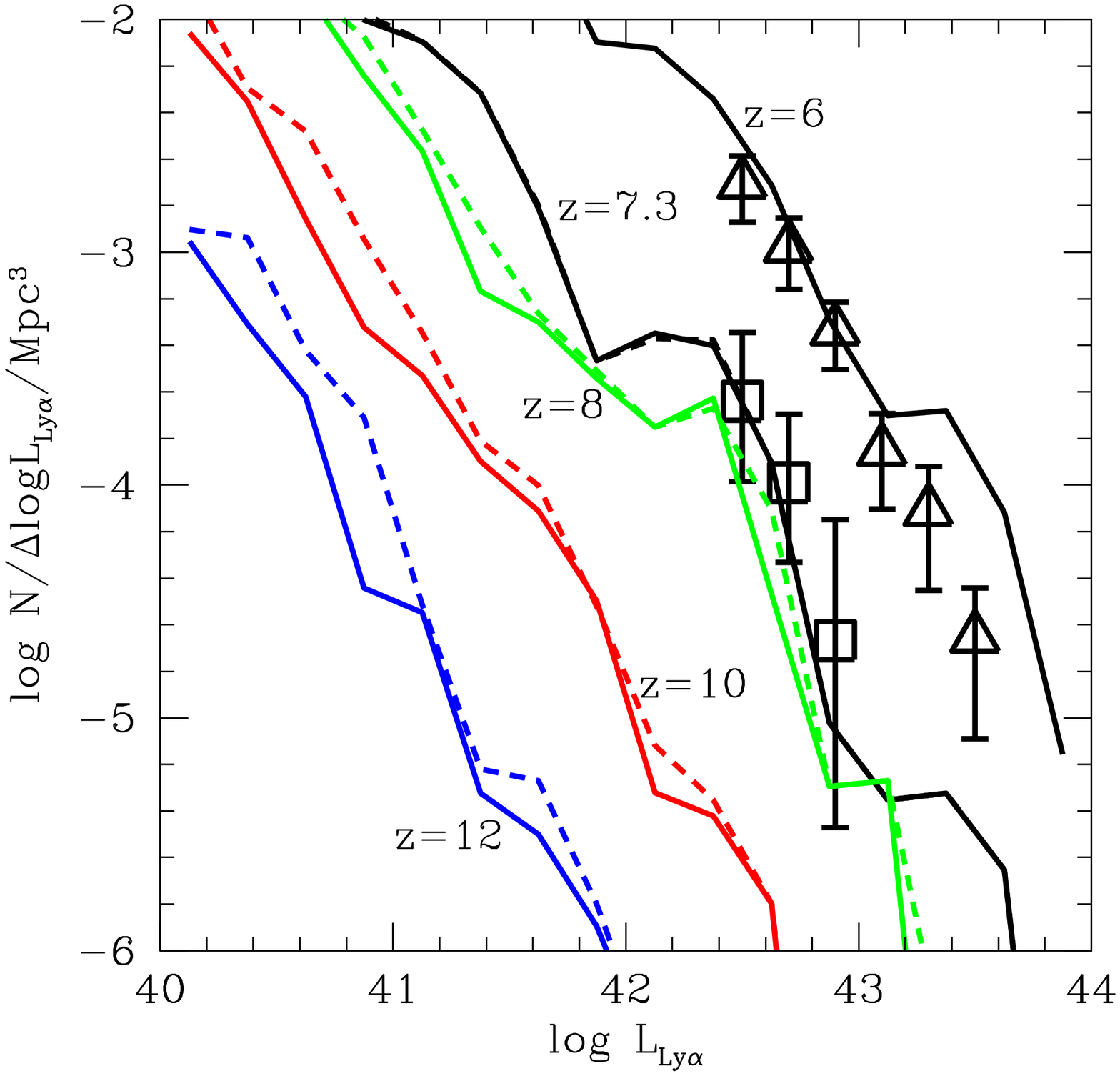}
\caption{
$\lya$ luminosity functions at $z=6, 7.3, 8, 10$ and $12$.
Open triangles and squares show observed LAEs at $z=5.7$ \citep{Ouchi08} and $7.3$ \citep{Konno14}.
Solid and dash lines represent the LFs based on the cloud and shell models. 
Solid line at $z=6$ shows the LF without the IGM attenuation. 
}
\label{fig:LFlae_z}
\end{center}
\end{figure}

Next we estimate the redshift evolution of LF based on the model A with
$V_{\rm out} = 180$ (cloud model) and $130~\rm \kms$ (shell model).  Figure~\ref{fig:LFlae_z} shows the
modeled LFs at $z=7, 8, 10$ and $12$.  Note that, here we use the
same line profile for all halos and redshift.  At higher redshifts,
typical SFR is smaller due to lower halo mass and halo growth rate.  In
addition, as redshift increases, typical size of H{\sc ii} bubbles
decreases, resulting in the lower IGM transmission.  As a result, the LF
rapidly shifts to the fainter side at higher redshifts.

The LFs of the shell models at faint-end are somewhat larger than the cloud model. 
As explained above, the bump at longer wavelength due to the back scattering effect
causes the higher IGM transmission, resulting in the higher LFs.
We also compare our modeled LF at $z=6$ with the observed LF at $z=5.7$ by \citet{Ouchi08}.
At this redshift, the cosmic reionization is thought to be completed, therefore
the IGM transmission is likely to be $\sim 100~\%$. 
Black solid line represents the LF at $z=6$ without the IGM attenuation. 
The modeled LF nicely matches the observation at $\La < 10^{43}~\rm \ergs$, but
somewhat larger at $\La > 10^{43}~\rm \ergs$. 
Bright LAEs are likely to reside in massive halos. 
As halos grow, interstellar gas could be dust enriched via type-II supernovae, 
resulting in the decrease of $\fesca$ \citep[e.g.,][]{Yajima14c}. 
The lower $\fesca$ might explain the lower LF at the bright end. 
In addition, \citet{Mesinger10} suggested that the IGM was not completely ionized even at $z \sim 5-6$. 
The residual neutral IGM could decrease number of observed LAEs.


\subsection{Relation between Ly$\alpha$ luminosity and stellar mass}
\label{sec:lyamst}

\begin{figure}
\begin{center}
\includegraphics[scale=0.42]{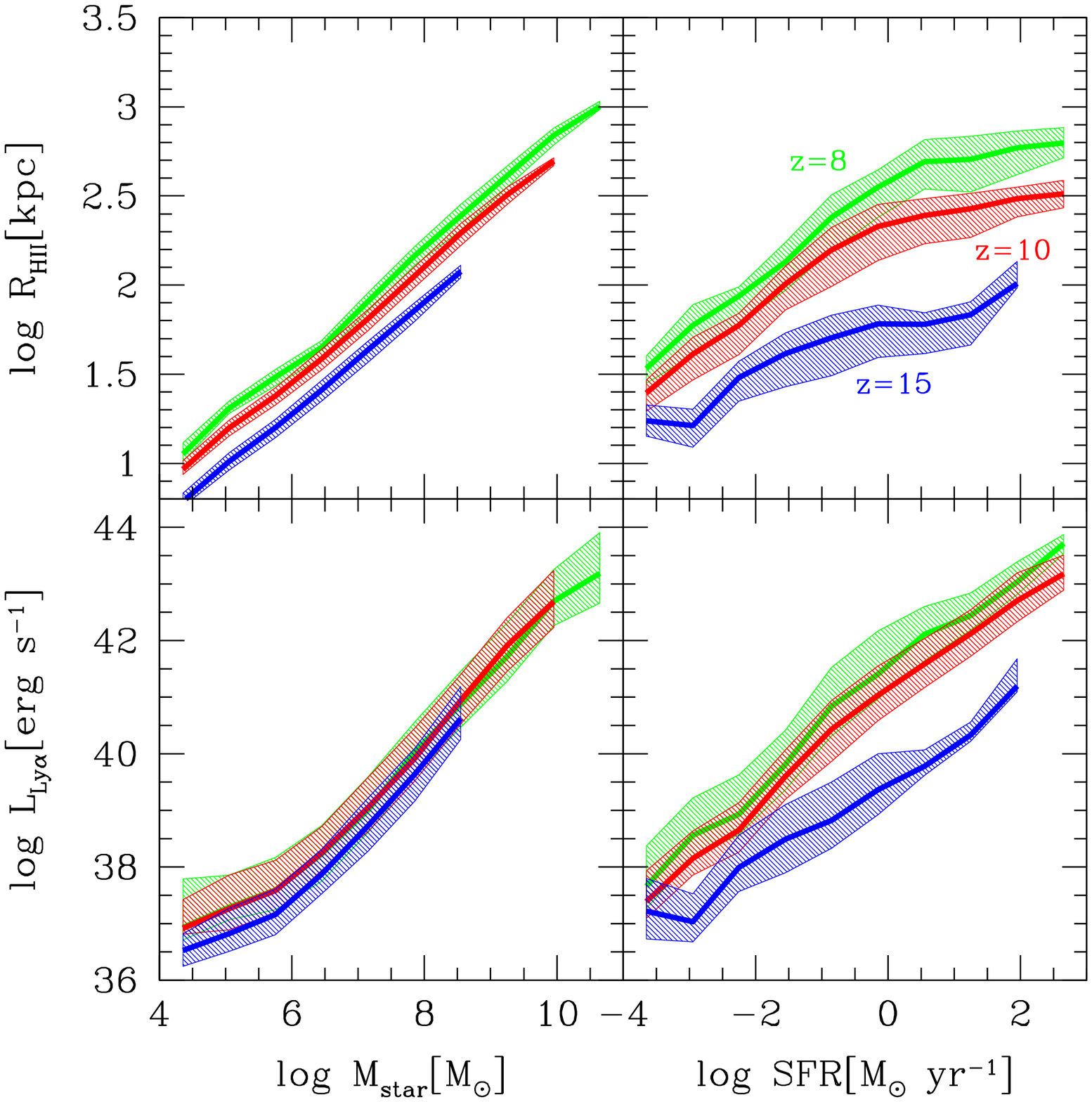}
\caption{
{Upper panel:} Sizes of H{\sc ii} bubbles as a function of stellar mass and SFR. 
Green, red and blue lines represent median values to galaxies at $z=8, 10$ and $15$, respectively. 
The shades show quartiles at each bin.
{Lower panel:} $\lya$ luminosity using the cloud model as a function of stellar mass and SFR.
}
\label{fig:Lyashade}
\end{center}
\end{figure}

Figure~\ref{fig:Lyashade} shows the sizes of H{\sc ii} bubbles and
$\lya$ luminosities considering the IGM transmission.  $\lya$ properties
are calculated by using the $\lya$ profile of model A of the cloud model.  The shades
represent the range of $25\% - 75\%$ in the sample.  We see that
$\rhii$ tightly correlate with stellar mass, as $\rhii \propto
\Mstar^{1/3}$, while the relation with SFR shows a large dispersion.

SFR rapidly increase by major merger.  However, 
$\rhii$ is not so sensitive to the short-time fluctuation of SFR because
of the longer time-scale for reaching the ionization equilibrium state.
As a result, the relation between $\rhii$ and SFR shows the
large dispersion.  High-redshift galaxies have been
observed as so-called Lyman-break galaxies \citep[LBGs:][]{Bouwens12}, via the Lyman-break technique so far.  
Our results indicate LBGs at $z
\gtrsim 7$ with similar UV brightness can have different $\lya$
fluxes due to the scatter of IGM transmission.
At $z=15$, our sample is limited by the stellar mass of
galaxies $\Mstar \sim 10^{9}~\Msun$.  In our
model, since the stellar mass is simply proportional to halo mass as
$\Mstar \sim 3.3\times10^{-3}~ \Mh$, this 
means that there is no progenitors with $\Mh \gtrsim 3 \times
10^{11}~\Msun$ at $z = 15$ in our halo sample, which
is constructed to have the mass range from $10^{9}$ to $10^{13}~\Msun$
at $z=6$.

In contrast to the SFR-$\rhii$ relation, both $\lya$ luminosity and $\rhii$
tightly correlate with $\Mstar$.
The $\La$-$\Mstar$ relation does not change with redshift significantly,
while the $\rhii$-$\Mstar$ becomes smaller as redshift increases.   
This is because Hubble constant (i.e., expanding velocity of
IGM) becomes large at higher redshift.  Therefore,
although $\rhii$ decreases as redshift increases due to
higher IGM density, the IGM transmission does not decreases
significantly.  Thus, $\lya$ luminosity does not depend sensitively
on redshift.
Note that, however, we have not taken the overlaps of H{\sc ii} bubbles into account. 
When galaxies are clustered,
giant H{\sc ii} bubbles can form by the overlaps of individual H{\sc ii} bubbles. 
This can cause scatters in the relations between $\Mstar$ and $\La$ or $\rhii$. 

The detection sensitivity of recent observations of LAEs at $z \sim 7 -
8$ was corresponding to $\lya$ luminosity of $\gtrsim 3\times10^{42}~\rm
\ergs$ \citep{Ono12, Shibuya12, Finkelstein13, Vanzella11, Konno14, Zitrin15}.  
In our model, median and minimum stellar masses producing $\La \sim 3 \times 10^{42}~\rm erg~s^{-1}$ at $z=7.3$ are $6.5\times10^{9}$ and $1.5\times10^{8}~\Msun$, respectively. Therefore, by considering the relation $\Mstar \sim 3.3\times10^{-3}~ \Mh$, we suggest that the observed LAEs at $z = 7.3$ should be hosted in halos with $\Mh \ge 4.6 \times 10^{10}~\Msun$, and the median halo mass is $1.9\times10^{12}~\Msun$.


\subsection{Redshift evolution of number density of observable LAEs}
\begin{figure*}
\begin{center}
\includegraphics[scale=0.6]{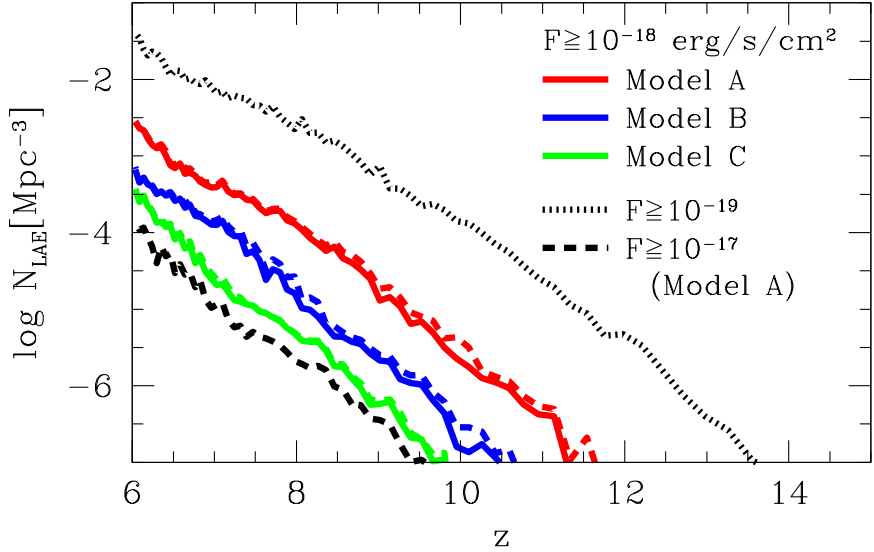}
\caption{ Number density of LAEs as a function of redshift.  
Red, blue
and green lines show the number densities of LAEs with
$F_{\lya} \ge 10^{-18}~\rm erg \, cm^{-2}\, s^{-1}$
obtained assuming $\lya$ line profiles of model A, B and
C, respectively.  
Colored solid and dash lines represent the number density using the line profiles of the 
cloud and shell models. 
Black dash and dot lines are the number density of
LAEs with $F_{\lya} \ge 10^{-17}~\rm erg \, cm^{-2}\,
s^{-1}$ and $F_{\lya} \ge 10^{-19}~\rm erg \, cm^{-2}\, s^{-1}$
obtained assuming the $\lya$ line profile of model A of the cloud model.  
}
\label{fig:nlae}
\end{center}
\end{figure*}

So far the $\lya$ line has been used as the most strong tool to confirm
the redshift of distant galaxy candidates \citep[e.g.,][]{Iye06, Finkelstein13,
Zitrin15}.  However, it is widely
thought that LAEs at $z > 9$ are difficult to be detected because of the
IGM opacity.  Here, we estimate the number density of LAEs ($\nlae$)
with higher $\lya$ flux than specific thresholds.  Figure~\ref{fig:nlae}
shows the number density of LAEs with $F_{\rm \lya} \ge 10^{-17},
10^{-18}$ and $10^{-19}~\rm erg~s^{-1}~cm^{-2}$.  The detection limits
of current observations with a reasonable integration time are $\sim
10^{-17}~\rm \ergscm$ \citep[e.g.,][]{Shibuya12}.  
As explained
in Sec.~3.2, the number density of bright LAEs monotonically decreases
with increasing redshift.  Given the detection limits of
$10^{-17}~\rm \ergscm$, wide field surveys of $~ 100^{3}~\rm Mpc^{3}$
are able to detect LAEs up to $z \sim 8.5$.  This is consistent with
recent observed LAEs at $z \lesssim 9$ \citep{Finkelstein13, Oesch15, Zitrin15}.  The LAEs with $F_{\rm \lya} \ge
10^{-17}$ at $z \sim 10$ are quite rare, with $n_{\rm LAE} \sim
1-2~\rm Gpc^{-3}$.

Spectroscopies of next
generation telescopes, e.g., JWST, are supposed
to achieve the sensitivity of $\sim 10^{-18}~\rm \ergscm$ with a
reasonable integration time.  If galaxies at $z \sim 10$ have outflowing
gas with $v \gtrsim 100~\rm km~s^{-1}$, the number density of LAEs with
$\Flya \ge 10^{-18}~\rm \ergscm$ at $z \sim 10$ is $\sim$ a few $\times
10^{-6}~\rm Mpc^{-3}$.  As shown in Figure~\ref{fig:Lyashade}, bright
LAEs are hosted in massive halos.  The median halo and stellar mass of
LAEs with $\Flya \ge 10^{-18}~\rm \ergscm$ at $z=10$ are $1.1 \times
10^{12}$ and $3.5 \times 10^{9}~\Msun$, respectively.  It was suggested
that the observed LBG at $z=11.1$, GNz11, had the stellar mass of $\sim
10^{9}~\Msun$ \citep{Oesch16}, which is corresponding to $\Flya \sim 0.2 \times 10^{-18}~\rm \ergscm$ in our model.  
Hence, it will be challenging to detect $\lya$
flux from GNz11 even by future spectroscopies with the line sensitivity
of $\Flya \sim 10^{-18}~\rm \ergscm$.

Different line profile models predict different
number densities of observable LAEs.  
The IGM transmission becomes more sensitive to the intrinsic $\lya$ line profile models,
since the typical size of H{\sc ii} bubbles gets smaller with increasing redshift.
As a result, the difference of $\nlae$ among model A, B and C becomes larger
at higher redshift, and more than order unity at $z \sim 10$.  Future
observation would also allow us to discriminate
intrinsic line profiles, which in turn provide
information about H{\sc i} column density and outflow velocity, by
comparing the observed number density of LAEs with the theoretical
models.
On the other hand, it is difficult to distinguish the cloud and shell models from the number density alone as shown in the figure.

Even next generation telescopes, e.g., GMT, E-ELT, TMT, will be
difficult to have the sensitivity of $\sim 10^{-19}~\rm \ergscm$.
However, if future telescopes somehow achieve such
a high sensitivity, wide field surveys of $\sim 100^{3}~\rm Mpc^{3}$ would be able to reach LAEs at $z \sim 12$.

In this work, we consider isolated H{\sc ii}
bubbles associated with individual LAEs in the estimation of IGM
transmission.  However, at lower redshift, H{\sc ii} bubbles can be
overlapped each other \citep[e.g.,][]{Iliev12, Hasegawa13, Ocvirk16}.  The overlapped H{\sc ii} bubbles can enhance the
IGM transmission.  This effect will be investigated in Hasegawa et
al. (in preparation) by combining large-scale $N$-body with small scale
radiative-hydrodynamics simulations.

\begin{table}
\begin{center}
 \caption{Model parameters}
\begin{tabular}{ccccccc}
\hline
Model &  $N_{\rm HI}/{\rm cm^{2}}$ & $V_{\rm out}/{\rm km\,s^{-1}}$ (cloud) & $V_{\rm out}/{\rm km\,s^{-1}}$ (shell)\\
\hline
A      & $2\times10^{20}$   & 180  & 130\\
B      & $2\times 10^{19}$   & 190  & 60\\
C     &  $2\times 10^{18}$   & 110 & 40 \\
\hline
  & $\fescion=0.2$ & \cr
  & $\fescuv=0.6$ & \cr
  & $\fesca=0.6$ & \cr
\hline
\end{tabular}
\begin{flushleft}
NOTES.\textemdash For each H{\sc i} column density, the outflow
velocity is chosen to reproduce the observed luminosity function of LAE
at $z = 7.3$ \citep{Konno14}.  
$V_{\rm out}$ (cloud) and $V_{\rm out}$ (shell)
represent best-fitted outflow velocities in the expanding cloud and shell models.
Escape fractions of ionizing,
UV and $\lya$ photons are same for all models.
\end{flushleft}
\label{table:model}
\end{center}
\end{table}

%
%

\section{Discussion} 
\label{sec:discussion}
\subsection{Impact of overlap of H{\sc ii} bubbles on $\lya$ radiation properties}
So far, we consider only isolated H{\sc ii} bubbles around LAEs. 
However, due to the clustering of galaxies, H{\sc ii} bubbles overlap each other \citep[e.g.,][]{McQuinn07a, Zahn11, Iliev12}. 
This leads to the expansion of the size of H{\sc ii} bubbles, resulting in the increase of IGM transmission. 
Recently, \citet{Castellano16} observed two LAEs in a galaxy overdensity region at $z \sim 7$ that might imply the clustering galaxies made a giant H{\sc ii} bubble. 
Here, considering the overlap effect, we artificially expand sizes of all H{\sc ii} bubbles associated with galaxies as follows:
\begin{equation}
R_{\rm HII}' = f_{\rm boost}^{\rm HII} R_{\rm HII},
\end{equation}
where $f_{\rm boost}^{\rm HII}$ is the boost factor to expand the original size of H{\sc ii} bubbles. 
For example, the condition of $f_{\rm boost}^{\rm HII}=2$ roughly considers that $\sim 8$ galaxies with similar luminosities distribute in an overlapped H{\sc ii} bubble, 
since the size of the H{\sc ii} bubbles is proportional to the power of one-third of luminosities.
In practice, $f_{\rm boost}^{\rm HII}$ can depend on original sizes of H{\sc ii} bubbles. 
Small H{\sc ii} bubbles associated with low-mass galaxies near massive ones
can be merged into giant H{\sc ii} bubbles, resulting in high $f_{\rm boost}^{\rm HII}$. 
On the other hand, the overlap between giant H{\sc ii} bubbles is not frequent, leading to low $f_{\rm boost}^{\rm HII}$. 

\begin{figure}
\begin{center}
\includegraphics[scale=0.43]{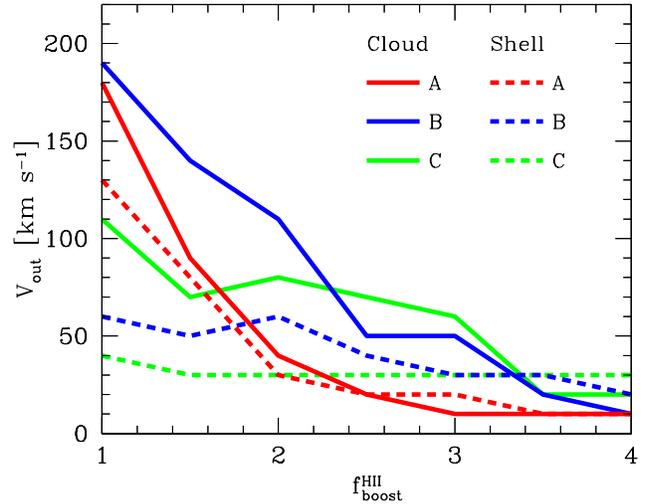}
\caption{
Outflow velocities matching the observed luminosity function 
as a function of the boost factor of H{\sc ii} bubble size. 
The IGM transmission is calculated to H{\sc ii} bubble extended artificially, 
$r'_{\rm HII} = f_{\rm boost}^{\rm HII} \times r_{\rm HII}$.
Solid and dash lines represent the expanding cloud and shell models, respectively.
Different colors indicate different H{\sc i} column densities in the calculations of line profiles.
}
\label{fig:vboost}
\end{center}
\end{figure}

As the size of H{\sc ii} bubbles increases by the overlap effect, even $\lya$ photons with the line center frequency can penetrate IGM. This is likely to affect the outflow velocity to reproduce the observation.  
Figure~\ref{fig:vboost} represents the best-fit outflow-velocities reproducing the observed LF at $z=7.3$ as a function of $\fboost$.
Due to the expansion of H{\sc ii} bubbles, the modeled LFs shift to brighter side, i.e., right side along x-axis in Figure~\ref{fig:LFlae}. 
Therefore, as $\fboost$ increases, the best-fit outflow-velocities monotonically decrease.  
All models show that the velocities become smaller than $\sim 50~\rm km~s^{-1}$ at $\fboost \gtrsim 3$.
However, in the range of $\fboost=1 - 4$, the velocities do not go below $0~\rm km~s^{-1}$, i.e., 
the LFs with $V=0~\rm km~s^{-1}$ do not exceed the observed LF. 
Therefore, if typical value of $\fboost$ is smaller than 4, LAEs are likely to have outflowing gas. 

The two-point correlation functions of LAEs derived from recent LAE surveys at $z=6.6$ showed the typical correlation 
length $r_{0}$ was $\sim 3~\rm Mpc$ in comoving scale and corresponding halo mass was $\sim 10^{11}~\Msun$ \citep{Ouchi10, Ouchi17}. The correlation length of LAEs does not change with redshift significantly \citep{Ouchi10}. 
In our model, the H{\sc ii} bubble size around halos of $\sim 10^{11}~\Msun$ is $\sim 260~\rm kpc$.
The fact that the H{\sc ii} bubble size is typically smaller than $r_{0}$ might indicate that $\fboost$ 
does not largely exceed unity. 
The detailed overlap effect should be investigated by future cosmological simulations. 

The velocities of model B and C of the cloud model does not change largely even if $\fboost$ increases. 
As seen in Figure~\ref{fig:inprof}, the shell model with a large outflow velocity produces the strong peaks at the line center frequency. 
Therefore, even if $\fboost$ is small, the velocities of the shell model do not increase largely. 

\subsection{Condition for galactic outflow from LAEs}

As shown in Section~\ref{sec:LFlae}, high-redshift LAEs are likely to have 
galactic wind with $V_{\rm out} \gtrsim 100~\rm \kms$.
Here we roughly derive the condition for making gas outflow with $V_{\rm
out} \gtrsim 100~\rm \kms$ in a spherical gas cloud model.
Supernova (SN) feedback can
be responsible for causing strong outflow in high-redshift low-mass
galaxies \citep[e.g.,][]{Kimm14, Kimm15}.  In the assumption that
supernovae give feedback to all gas uniformly, we estimate the total
kinetic energy of outflowing gas as
\begin{equation}
\begin{split}
 \frac{1}{2} \Mgas V_{\rm out}^{2}  \sim f_{\rm conv} e_{\rm SN} N_{\rm SN}
- \frac{G \Mh \Mgas}{R_{\rm vir}}.
\end{split}
\end{equation}
Here $\Mgas=(1-\epsilon)\Mgas^{0}$ is gas mass after star
formation, $\Mstar=\epsilon\Mgas^{0}$ is stellar mass, $\Mgas^{0}$
is initial gas mass, $\epsilon \equiv \Mstar / \Mgas^{0}$ is a star
formation efficiency with respect to the initial gas mass, 
$f_{\rm conv}$ is the conversion efficiency from total supernova energy to kinetic one of gas,
$e_{\rm SN} \sim 2 \times 10^{51}~\rm erg$ is the released energy for each supernova  \citep[e.g.,][]{Hamuy03},
and $N_{\rm SN}$ is the number of supernova.
For Salpeter-like IMF, $N_{\rm SN} \sim 1\times 10^{-2} \frac{\Mstar}{\Msun} = 1\times 10^{-2} \left( \frac{\epsilon}{1-\epsilon}\right) \frac{\Mgas}{\Msun}$.
By using the star formation efficiency $\epsilon$, we simply estimate $V_{\rm out}$ as
\begin{equation}
\sqrt{V_{\rm out}^{2} + v_{\rm esc}^{2}} \sim 1.4\times10^{3}~{\rm km~s^{-1}}~ \left( \frac{\epsilon} {1-\epsilon} \right)^{\frac{1}{2}} f_{\rm conv}^{\frac{1}{2}}, 
\label{eq:vout}
\end{equation}
where $v_{\rm esc} \equiv \sqrt{\frac{2G\Mh}{R_{\rm vir}}}$ is the escape velocity of a halo. 

Next we obtain $f_{\rm conv}$ as a function of $\epsilon$. 
Recently \citet{Kim15} showed the final momentum produced by a single SN with the energy of $10^{51}~\rm erg$ as follows:
$p = 2.8 \times 10^{5} ~{\rm \kms~\Msun} \left( \frac{\nh}{1~\rm cm^{-3}} \right)^{-0.17}$ \citep[see also,][]{Cioffi88, Thornton98}.
In this work, we ignore the weak density dependence and assume $\nh \sim 1~\rm cm^{-3}$.
Since the final momentum is almost linearly proportional to the injected SN energy \citep{Cioffi88, Kimm14}, 
we approximate the total momentum produced by multiple SNe as follows:
$P = p \left( \frac{e_{\rm SN} N_{\rm SN}}{10^{51}~\rm erg} \right)$, where $e_{\rm SN} N_{\rm SN}$ is the total energy of supernovae. 
Therefore, using $E = \frac{P^{2}}{2\Mgas}$, we derive $f_{\rm conv} = \frac{E}{E_{\rm SN}} = 16 \left( \frac{\epsilon}{1-\epsilon}\right)$.
Note that $f_{\rm conv}$ can not exceed unity according to the energy conservation. 
For that reason, we set $f_{\rm conv}=1$ at $\epsilon \ge 0.06$ because the above expression gives $f_{\rm conv} > 1$. 
Thus, Equation~\ref{eq:vout} is written by using $\epsilon$ as follows:
\begin{equation}
\sqrt{V_{\rm out}^{2} + v_{\rm esc}^{2}} \sim \begin{cases}
5.5 \times10^{3}~{\rm km~s^{-1}}~ \left( \frac{\epsilon} {1-\epsilon} \right)\\ 
~~~~~~~~~~~~~~~~~~~~~~~~~~~~~~~~~{\rm if}~~ \epsilon < 0.06\\
1.4 \times10^{3}~{\rm km~s^{-1}}~ \left( \frac{\epsilon} {1-\epsilon} \right)^{\frac{1}{2}} \\  
~~~~~~~~~~~~~~~~~~~~~~~~~~~~~~~~~{\rm if}~~ \epsilon \ge 0.06.
\end{cases}
\end{equation}
We find that the condition of $\epsilon \sim 0.04$ is required to cause the galactic outflow with $V_{\rm out}=180~\rm \kms$
from a halo with $\Mh=4.6 \times10^{10}~\Msun$ at $z=7.3$ which is minimum halo mass to produce the observable $\lya$ luminosity $\sim 3\times10^{42}~\rm erg~s^{-1}$. 
We can also convert the tuning parameter $\alpha$ in our star formation model to $\epsilon$,
as  $\epsilon \sim \alpha \Omega_{\rm M}/\Omega_{\rm b} \sim 0.02$.
This is roughly similar to the value estimated above. 
As halo mass increases, higher $\epsilon$ is required to cause galactic outflow.
Here we have estimated $\epsilon$ assuming all gas has same outflow velocity in the simple spherical cloud model. 
For example, in the case of disk galaxies, 
only a part of gas can be evacuated along the normal direction to galactic disk
as shown in numerical simulations \citep[e.g.,][]{Agertz11}. 
In this case, strong outflow can be caused even with smaller $\epsilon$
because piled gas mass can be lower. 

\subsection{Relation between Ly$\alpha$ luminosity and size of ionized bubble}

\begin{figure}
\begin{center}
\includegraphics[scale=0.4]{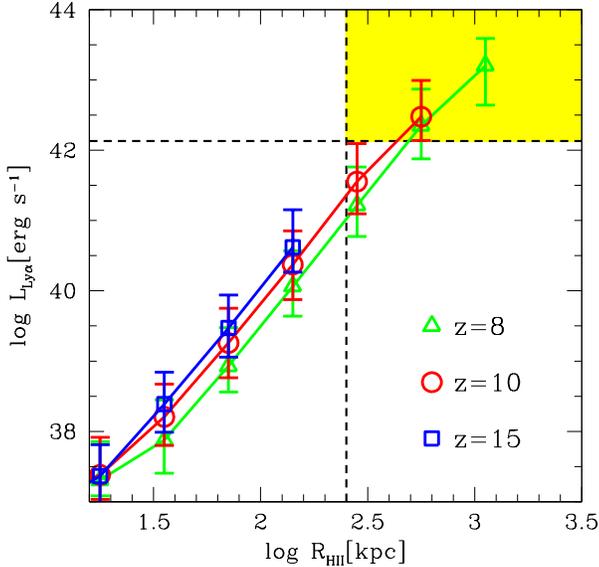}
\caption{
$\lya$ luminosity as a function of size of H{\sc ii} bubble. 
Triangle, circle and square symbols represent median values
at $z=8, 10$ and $15$, respectively. 
Error bars show quartiles. 
The $\lya$ line profile of model A is used.
Horizontal dash line shows $\lya$ flux of $10^{-18}~\rm erg\, cm^{-2}\, s^{-1}$ for $z=10$.
Vertical dash line represents the viewing angle of 1 arcmin for $z=10$.
}
\label{fig:LyaRhii}
\end{center}
\end{figure}

LAEs can be responsible for ionizing sources of cosmic reionization
\citep[e.g.,][]{Yajima09, Yajima11, Yajima14c}.  Figure~\ref{fig:LyaRhii} shows
$\Llya$ as a function of the size of associated H{\sc ii} bubble
$\rhii$.  The $\lya$ luminosity steeply increases with $\rhii$ due to higher IGM transmission for large H{\sc ii} bubbles. 
The relation between $\Llya$ and $\rhii$ does not significantly change with redshift.
IGM gas density increases with redshift as $\rho \propto (1+z)^{3}$,
resulting in lower IGM transmission at higher redshift. 
However, galaxies tend to have higher ionizing photon emissivity and
intrinsic $\lya$ luminosity for a fixed size of H{\sc ii} bubbles at
higher redshift.  The combination of these effects leads to the weak
redshift dependence in the $\La$-$\rhii$ relation.

The yellow shaded region represents
 the viewing angle of $\ge 1$ arcmin and the flux of $\ge 10^{-18}~\rm \ergscm$ for $z \sim 10$.
This region corresponds to the LAEs with
associated H{\sc ii} bubbles that are detectable both as LAEs and 
holes in 21-cm signal by future galaxy observations by JWST and 21-cm tomography
by SKA-2, respectively.  Future 21-cm observations will be able to
probe giant H{\sc ii} bubbles around bright LAEs with $\Flya \gtrsim
10^{-18}~\rm \ergscm$ at $z \sim 10$.
Note that, the overlap of H{\sc ii} bubbles is not considered in Figure~\ref{fig:LyaRhii}.
The effect of the overlap can shift our results to larger side of the H{\sc ii} bubble. 

The differential brightness temperature $\delta T_{\rm b}$ caused by
galaxies shows inner positive and outer negative ring-like
structure \citep[e.g.,][]{Chen04, Yajima14e}.  The
detailed structure depends on SED. If galaxies host X-ray sources like
AGNs, the positive region is extended due to partial photo-ionization
heating.  Therefore,
future 21-cm observations may also give us information about nature of X-ray and
UV sources in bright LAEs.

\subsection{Limitations in current models}
\label{sec:modellimit}

Here we have investigated properties of LAEs and ionized bubbles at the era of reionization
based on simple analytical models. 
Our current model is not able to consider following important aspects.

(1) redshift evolution of parameters: In this work, we have assumed the escape fraction of $\lya$ photons and 
the shape of $\lya$ line profile did not change with redshift. 
These quantities depend on physical properties of interstellar medium
that can change with redshift \citep[e.g.,][]{Mao07, Yajima14c}. 

(2) overlap of ionized bubbles:
As ionized bubbles grow, they overlap each other, leading to formation of giant H{\sc ii} bubbles. 
This increases the IGM transmission of $\lya$ photons with the frequency near the line center. 

(3) $\lya$ surface brightness: Multiple scattering processes of $\lya$ photons in galaxies can make extended $\lya$ surface brightness distribution. Recent observations indicated star-forming galaxies had extended $\lya$ haloes \citep[e.g.,][]{Steidel11, Momose14, Momose16}. Therefore, some fraction of $\lya$ flux could be lost in the observations of LAEs. 
Understanding the $\lya$ surface brightness distribution and the flux loss are quite important in liking between theoretical models and observations.
However, the fraction of lost $\lya$ flux sensitively depends on the sensitivities of the observations, and it is uncertain now.  In this work, we have assumed that all $\lya$ photons escaped from galaxies can be 
observed, if they are not scatted by IGM. 
This assumption can be reasonable when sizes of galaxies are not much larger than 
angular resolutions of $\lya$ observations. 
Recently \citet{Mas-Ribas17} showed JWST will be able to reach the sensitivity
$\sim 10^{-20}~\rm erg~cm^{-2}~s^{-1}~arcsec^{-2}$ for extended emission sources. 
Here we roughly estimate the surface brightness as
$S_{\lya} \sim \frac{F_{\lya}}{\pi r_{\rm edge}^{2}}$.
For detections of bright LAEs of $F \sim 10^{-18}~\rm erg~cm^{-2}~s^{-1}$, the source size should be smaller than $\sim 6~\rm arcsec$, corresponding to $r_{\rm edge} \sim 30~\rm kpc$ at $z \sim 7$. 
This is comparable with the virial radius of halos with masses of $10^{11} - 10^{12}~\Msun$. 
Therefore the flux loss due to the faint surface brightness is unlikely to be significant. 
In addition, \citet{Momose14} suggested that typical scale lengths of $\lya$ halos of observed LAEs at $z \sim 2-6$ were $\sim 5-10~\rm kpc$ and do not change with redshift significantly.  
Note that, however, the surface brightness sensitively depends on complicated distributions of gas and velocity fields \citep[e.g.,][]{Yajima15}.
In reality, the surface brightness distribution should be considered in the estimation of the flux loss for extended sources, which is beyond the scope in this paper.  
We will study the detectability of distant LAEs with the calculation of surface brightness
by combining cosmological hydrodynamics simulations and $\lya$ radiative transfer calculations in future work.
 
%
%

\section{Summary}
\label{sec:summary}

We present models of $\lya$ emitting galaxies (LAEs) with IGM transmission considered at the era of
reionization.  Based on halo merger trees and a simple star formation
model, we estimate cosmic star formation and cosmic reionization
history. Our model uses 5000 realizations of halo merger trees with the
halo mass range from $\Mh=10^{9}$ to $10^{13}~\Msun$ at $z=6$.  As a
result, our model reproduces the observed cosmic star formation
densities, stellar mass densities, luminosity functions of Lyman-break
galaxies with a tuning parameter, $\alpha (\equiv SFR /
\frac{d\Mh}{dt}) =3.3 \times 10^{-3}$.  Our
modeled star formation history, with an escape fraction of ionizing
photons $\fescion = 0.2$, also provides a
cosmic reionization history consistent with the
Thomson scattering optical depth indicated by Plank
(2016).

Based on the above parameters, we model LAEs and H{\sc ii} bubbles using
individual halo merger trees.  Our models show the distribution function of
the size of H{\sc ii} bubble, and indicate giant H{\sc ii} bubbles
associated with bright LAEs at $z \lesssim 12$ can
be probed by future 21-cm observation using SKA.  We find that $\lya$
flux is tightly related with stellar/halo mass,
while there is a large dispersion in the relation between $\lya$ flux
and SFR.  By comparing our models with the observed luminosity function
of LAEs at $z=7.3$ by \citet{Konno14}, we indicate that LAEs are
likely to have the H{\sc i} column density of $N_{\rm HI} \gtrsim
10^{20}~\rm cm^{-2}$ and the outflowing gas with $V_{\rm out} \gtrsim
100~\rm \kms$.  Using these parameters, we predict that future wide deep
survey can detect LAEs at $z \sim 10$ with $n_{\rm LAE} \sim {\rm a~few~}\times 10^{-6} ~\rm Mpc^{-3}$
and $\sim 1 \times 10^{-4} ~\rm Mpc^{-3}$ for the flux
sensitivity of $10^{-18}~\rm erg\, cm^{-2}\, s^{-1}$ and
$10^{-19}~\rm erg\, cm^{-2}\, s^{-1}$, respectively.  Our models predict next
generation telescopes, JWST, E-ELT or TMT would be able to observe
galaxies at $z \sim 10$ via the detections of $\lya$ lines.
 
By combining future galaxy observations and 21-cm tomography with SKA-2, we will
be able to detect both LAEs with $\La \gtrsim 10^{42}~\ergs$
and their associated giant H{\sc ii} bubbles with $\rhii \gtrsim
250~\rm kpc$.  We suggest that a clear spatial
anti-correlation between galaxies and 21-cm emission will be detected by
focusing on such bright LAEs.
Here we consider individual H{\sc ii} bubbles associated LAEs alone. 
However, when galaxies distribute in clustering regions, 
the H{\sc ii} bubbles overlap each other, resulting in the formation of giant H{\sc ii} bubbles. 
Therefore, this can cause a large dispersion in the relation between the bubble size and $\lya$ luminosity. 

In this work, we assume that the H{\sc i} column density and outflow velocity do not change significantly 
with redshift. 
On the other hand, \citet{Konno16} suggested that 
H{\sc i} column density should decrease as the redshift increases
over the redshift range $z \sim 2 - 6$ in order to reproduce the redshift evolution of observed $\lya$ escape fraction.  
They suggested that LAEs at $z \sim 6$ should have the H{\sc i} column density of $\sim 10^{18}~\rm cm^{-2}$
to obtain the high $\lya$ escape fraction. 
However, the relation between the H{\sc i} column density and the escape fraction can be changed due to physical properties of ISM. 
Recently \citet{Gronke16} have introduced some detailed properties of ISM into their expanding shell models, e.g., clumpiness. 
The porous ISM can reproduce high escape fraction even for high H{\sc i} column density. 
This can alleviate the discrepancy between our models and \citet{Konno16}. 
Note that, in our model, the high H{\sc i} column density has been favored to obtain reasonable IGM transmissions 
for reproducing the observed luminosity function. 
However, when the overlaps of H{\sc ii} bubbles are considered, even somewhat lower H{\sc i} column density can be allowed. 
Lower H{\sc i} column density leads to the smaller shift of the peak frequency of $\lya$ line profile, resulting in lower IGM transmission. 
Therefore, the number density of observable LAEs at $z \gg 7$ becomes smaller than the above estimation,  if the H{\sc i} column density decreases. 

In addition, the flux loss due to limited sensitivities in observations has not been taken into account 
in our current models. 
Even the high-sensitivity of JWST can lose some fraction of total flux
if high-$z$ LAEs are very extended due to $\lya$ scattering in ISM. 

Thus our estimation of detectability for LAEs at $z \gtrsim 10$ might be somewhat optimistic. 
We will investigate the $\lya$ line profiles and the surface brightness distribution by combining 
cosmological hydrodynamics simulations and $\lya$ radiative transfer calculations 
in our future work. 

%
%
\section*{Acknowledgments}
We thank Akira Konno and Masami Ouchi for providing us their recent
observational data.  We are grateful to Masayuki Umemura, Akio Inoue,
Yoshiaki Ono, Takatoshi Shibuya, Daisuke Nakauchi and Sadegh Khochfar for valuable discussion and
comments.  The numerical simulations were performed on the computer
cluster, {\tt Draco}, at Frontier Research Institute for
Interdisciplinary Sciences of Tohoku University.  This work is supported
in part by MEXT/JSPS KAKENHI Grant Number 17H04827 (HY) and 15J03873 (KS).

%
%

\label{lastpage}

\end{document}